\documentclass{article}

\usepackage{authblk}
\usepackage[left=2cm,right=2cm,top=2cm,bottom=2cm]{geometry}
\usepackage[utf8]{inputenc}    
\usepackage{graphicx}
\usepackage[comma,super,sort]{natbib}

\usepackage{color}

\usepackage[resetlabels,labeled]{multibib}
\newcites{Methods}{References}
\usepackage{caption}
\usepackage{amsmath}
\usepackage{amsfonts}
\usepackage{amssymb}
\usepackage{subfigure}
\usepackage{wrapfig}
\usepackage{dcolumn}
\usepackage{bm}
\usepackage{comment}
\usepackage{mathrsfs,accents,bigints}

\title{Magneto-active elastic shells with tunable buckling strength}

\author[1]{{Dong Yan}}
\author[1]{{Matteo Pezzulla}} 
\author[1,2]{{Lilian Cruveiller}}
\author[1]{{Arefeh Abbasi}}
\author[1]{{Pedro M. Reis\footnote{To whom correspondence should be sent: pedro.reis@epfl.ch}}} 

\affil[1]{Flexible Structures Laboratory, Institute of Mechanical Engineering, \protect\\ \'{E}cole Polytechnique F\'{e}d\'{e}rale de Lausanne (EPFL), Switzerland}
\affil[2]{\'{E}cole Polytechnique, Palaiseau, France}

\date{}

\begin{document}

\maketitle
\vspace{-20pt}
\begin{quote}
\textbf{
Shell buckling is central in many biological structures and advanced functional materials, even if, traditionally, this elastic instability has been regarded as a catastrophic phenomenon to be avoided for engineering structures. Either way, predicting critical buckling conditions remains a long-standing challenge. The subcritical nature of shell buckling imparts extreme sensitivity to material and geometric imperfections. Consequently, measured critical loads are inevitably lower than classic theoretical predictions. Here, we present a robust mechanism to dynamically tune the buckling strength of shells, exploiting the coupling between mechanics and magnetism. Our experiments on pressurized spherical shells made of a magnetorheological elastomer demonstrate the tunability of their buckling pressure via magnetic actuation. We develop a theoretical model for thin magnetic elastic shells, which rationalizes the underlying mechanism, in excellent agreement with experiments. A dimensionless magneto-elastic buckling number is recognized as the key governing parameter, combining the geometrical, mechanical, and magnetic properties of the system.} 
\end{quote}

Shells are curved thin structures that can withstand extreme loading conditions due to the interplay between bending and stretching deformation~\cite{Koiter_NonlinearBuckling1969,Niordson_ShellTheory1985} through the so-called `shell effect'~\cite{Hutchinson_EML2020}. Thin shells are an ubiquitous structural element in engineering~\cite{Hilburger_AIAA2012} and also widely observed in nature across length scales, from viruses~\cite{Lidmar_PRE2003}, and capsules~\cite{sacanna2010,Datta_PRL2012,Vian_SoftMatter2019}, to pollen grains~\cite{Katifori_PNAS2010}, and plants~\cite{Forterre_Nature2005}. While curvature is the key ingredient underlying the excellent mechanical performance of shells, it is also responsible for the catastrophic (subcritical) nature of their elastic instabilities~\cite{misbah2016}. Consequently, shells are high sensitivity to imperfections~\cite{Tsien_AeronauticalSciences1942,Koiter_NonlinearBuckling1969,Hutchinson_JAM1967,Carlson_ExpMech1967}. For over a century, the pressure buckling of a spherical shell has been a long-standing canonical problem of elastic (in)stability~\cite{Koiter_NonlinearBuckling1969,Niordson_ShellTheory1985}. When the in-out pressure differential across a shell exceeds a threshold value, the shell loses its load-carrying capacity~\cite{Koiter_NonlinearBuckling1969}. The ensuing collapse is unpredictable, occurring at loads that are significantly lower than classic theoretical predictions~\cite{Carlson_ExpMech1967}. Consequently, measuring and predicting the critical buckling pressure has proven to be nontrivial due to the high imperfection sensitivity. 

In 1915, Zoelly~\cite{Zoelly_PhDThesis1915} derived
the theoretical prediction for the critical buckling pressure of a perfect spherical shell, of radius $R$ and thickness $h$, loaded under a uniform pressure $p$:
\begin{equation}
p_\mathrm{c}=\frac{2E}{\sqrt{3(1-\nu^2)}}\left(\frac{R}{h}\right)^{-2}\,, 
\label{eq:classic_pc}
\end{equation}
where $E$ and $\nu$ are the Young's modulus and Poisson's ratio of the material, respectively. Notwithstanding this classic result, given the high imperfection sensitivity, buckling pressures measured in experiments, $p_\mathrm{max}$, have long been found~\cite{Tsien_AeronauticalSciences1942,Hutchinson_JAM1967,Carlson_ExpMech1967,Koiter_NonlinearBuckling1969} to be much lower than the prediction from Eq.~(\ref{eq:classic_pc}). This discrepancy generated a long debate in the shell mechanics community that lasted for nearly four decades until the disagreement between theory and experiments was finally attributed to imperfections~\cite{Elishakoff2014}. Given this intrinsic mismatch, it became standard to define an empirical quantity, the so-called \textit{knockdown factor},
\begin{equation}
\kappa_\mathrm{d}=\frac{p_\mathrm{max}}{p_\mathrm{c}}\,,
\label{eq:kd}
\end{equation}
which is always smaller than unity, spreading a wide range~\cite{Carlson_ExpMech1967,Lee_JAM2016}: ${0.05} \le \kappa_\mathrm{d} \le {0.9}$. 

In engineering, significant efforts have focused on improving predictions for the knockdown factor and understanding how it is affected by imperfections~\cite{Tsien_AeronauticalSciences1942,Hutchinson_JAM1967,Carlson_ExpMech1967,Koiter_NonlinearBuckling1969}. A breakthrough came only recently, from experiments, when the combination of a rapid prototyping technique for spherical shells~\cite{Lee_NatCommun2016} and its adaptation to seed precisely designed defects led to a quantitative relationship between the knockdown factor and defect geometry~\cite{Lee_JAM2016}. This study demonstrated that if imperfections can be measured precisely, then the knockdown factor can be precisely predicted using an appropriate shell theory~\cite{Hutchinson_ProcMathPhysEngSci2016,Hutchinson_PhilosTransRoyalSocA2017,Jimenez_JAM2017,Ning_JMPS2018,Gerasimidis_JAM2018,Sieber_JAM2020,Yan_JMPS2020}, thus opening the door for less conservative designs of shell structures. In parallel, nondestructive probing methods have recently been proposed to experimentally access the stability landscape of shells~\cite{Virot_PRL2017,Marthelot_JAM2017,Thompson_IntJBifurcatChaos2017}, while accepting the inevitable presence of multiple imperfections to set the knockdown factor. Furthermore, bilayer shells undergoing differential swelling were recently shown to have a varying knockdown factor during a transient, demonstrating how the knockdown factor can be modified post-fabrication, albeit not reversibly on-demand~\cite{Lee_SoftMatter2019}. Still, to date, the knockdown factor is regarded as an intrinsic structural property dictated by imperfections imparted during fabrication or encoded at the design stage.

Here, we propose a robust mechanism to dynamically tune the buckling strength of shells (\textit{i.e.}, their knockdown factor), by coupling elastic deformation and magnetic actuation, to provide active control and liberate the predestination of imperfections. Our framework leverages recent advances in magneto-elastic mechanical systems -- including shape-programmable materials~\cite{Loukaides_IntJSmartNanoMater2014,Seffen_SmartMaterStruct2016,Lum_PNAS2016,Psarra_JMPS2019}, biomedical devices~\cite{Hu_Nature2018}, and soft robots~\cite{Kim_Nature2018,Kim_SciRobot2019,Kim_SciRobot2019,Gu_NatCommun2020} --, which we augment in the context of shell mechanics. Specifically, we manufacture spherical shells made of a hard-magnetic elastomer (hereon referred to as magnetic shells) and characterize their critical buckling pressure under an applied magnetic field. We demonstrate that the knockdown factor can be modified externally by adjusting the magnitude and polarity of the field. To rationalize our results, we develop a magnetic shell theory, providing a physical interpretation of the interplay between shell mechanics and magnetic fields. Furthermore, we uncover the magneto-elastic buckling number (a dimensionless quantity that combines the magnetic, elastic, and geometrical properties), which acts as the single governing parameter of the system.

\begin{figure}[h!]
\centering
\includegraphics[width=0.95\textwidth]{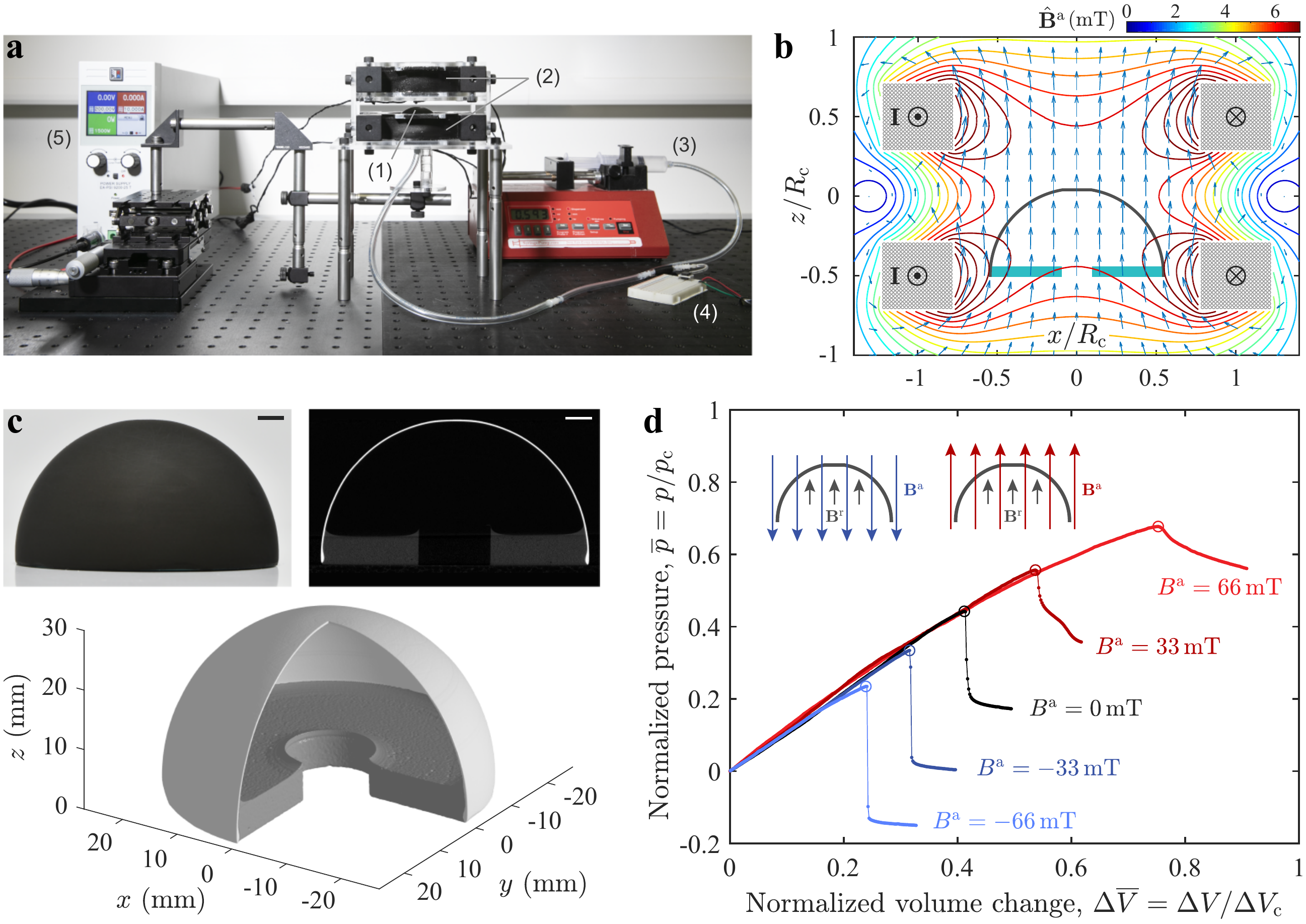}
\caption{\textbf{Buckling of a magnetic shell under combined pressure and magnetic loading.} \textbf{a}, Photograph of the experimental apparatus: A (1) magnetic shell is positioned in between (2) a set of Helmholtz coils and depressurized using  a pneumatic-loading system, comprising a (3) syringe pump  and (4) a pressure sensor . The Helmholtz coils are driven by the current $\mathbf{I}$ in the direction shown in \textbf{b}, output from a DC power supply (5). \textbf{b}, Computed field of the vector (arrow) and magnitude (contour line) of the magnetic flux density $\hat{\mathbf{B}}^\mathrm{a}$ generated by the coils, under current $I=1\,$A (Supplementary Note S2). The field is uniaxial and uniform in the central region with a flux density of ${\mathbf{B}}^\mathrm{a}={B}^\mathrm{a}\mathbf{e}_3$. Given axisymmetry, the field is represented in the $x$-$z$ plane ($y=0$). Coordinates $x$ and $z$ are normalized by the coil radius $R_\mathrm{c}=46.5\,$mm. \textbf{c}, Photograph of the representative imperfect shell specimen (top left). Image of its cross-section (top right) scanned through x-ray $\mu$CT and the reconstructed 3D view (bottom, one quarter was artificially hidden to aid with visualization). Scale bars are $5\,$mm. \textbf{d}, Loading curves of pressure, $\overline{p}=p/p_\mathrm{c}$, versus volume change, $\Delta \overline{V}=\Delta V/\Delta {V}_\mathrm{c}$, for the shell shown in \textbf{c}. The peak value represented by the open circle on each curve is the critical buckling pressure. The buckling test is performed at different levels of external flux density ${B}^\mathrm{a}$. Inset: Schematic of a shell subjected to a magnetic field, which is opposite (left) or parallel (right) to the shell magnetization.}
\label{fig:Fig1}
\end{figure}

\noindent \textbf{Experiments on tuning the knockdown factor of pressurized magnetic shells}\\
In our experiments, we position a hemispherical shell of radius, $R=25.4\,$mm, and thickness, $h=321.2\,\mu$m, in between a set of Helmholtz coils (Fig.~\ref{fig:Fig1}\textbf{a},\textbf{b}). These two coils impose a steady, uniaxial and uniform magnetic field (flux density vector ${\mathbf{B}}^\mathrm{a}={B}^\mathrm{a}\mathbf{e}_3$) on the shell, perpendicularly to its equatorial plane (see Methods and Supplementary Notes S1 and S2 for details). The shell is made of a magnetorheological elastomer (MRE), a composite of hard-magnetic NdPrFeB particles (volume fraction~$c=7\%$) and vinylpolysiloxane (VPS) polymer (see Methods). The full 3D configuration and the corresponding cross-section profile of the shell are visualized through X-ray micro-computed tomography ($\mu$CT,  100 Scanco Medical AG), a representative example of which is presented in Fig.~\ref{fig:Fig1}\textbf{c}. To measure the buckling strength, we depressurize the shell using a pneumatic-loading system under imposed-volume conditions and measure the associated pressure sustained by the shell (Methods and Supplementary Note S2). Figure~\ref{fig:Fig1}\textbf{d} presents the load-carrying behavior of the shell, characterized by the pressure ($p$) as a function of the volume change ($\Delta V$), both of which are normalized, respectively, by the classic buckling prediction, $p_\mathrm{c}$, and the corresponding volume change immediately prior to buckling~\cite{Hutchinson_ProcMathPhysEngSci2016,Hutchinson_PhilosTransRoyalSocA2017}, $\Delta{V}_\mathrm{c}={2\pi(1-\nu)R^2h}/{\sqrt{3(1-\nu^2)}}$. The onset of buckling corresponds to the maximum of each curve, $\overline{p}_{\mathrm{max}}={p}_{\mathrm{max}}/p_\mathrm{c}$, and the accompanying pressure drop indicates the loss of load-carrying capacity of the shell. 

\captionsetup[figure]{labelfont={bf},name={Fig.}}

In the absence of magnetic field (${B}^\mathrm{a}=0\,$mT), the representative shell shown in Fig.~\ref{fig:Fig1}\textbf{c} buckles at the dimensionless pressure level of~$\overline{p}_{\mathrm{max}}=0.44$. This value is far below the classical prediction of 1 (${p}_{\mathrm{max}}=p_\mathrm{c}$), the reason being that we intentionally seed a precisely engineered dimple-like defect at the pole during manufacturing (Methods and Supplementary Note S1), so as to consider the influence of imperfections in a controllable manner. Hence, the geometry of the shell deviates slightly from a perfect hemisphere by an amplitude of~$\delta$ over the region of polar angle~$0 \le \varphi \le \varphi_\circ$. The half angular width of the defect is then measured by $\varphi_\circ$, usually rescaled as~$\overline{\varphi}_\circ={\varphi_\circ}{(\sqrt{12(1-\nu^2)}R/h)^{\frac{1}{2}}}$~\cite{Lee_JAM2016, Hutchinson_ProcMathPhysEngSci2016,Jimenez_JAM2017,Yan_JMPS2020}. For this test shell, the defect geometry is characterized as~$\overline{\delta}=\delta/h=0.39$ and~$\varphi_\circ=11.7^\circ$ ($\overline{\varphi}_\circ=3.2$) using an optical profilometer (Methods and Supplementary Note S1). Although the introduced defect is too small to be seen with the naked eye, the buckling strength of the shell is dramatically reduced, by more than a half due to the high sensitivity to imperfections~\cite{Koiter_NonlinearBuckling1969,Hutchinson_ProcMathPhysEngSci2016,Lee_JAM2016}. 
 
We proceed by focusing on the effect of the magnetic field on the buckling instability. The magnetized shell possesses a residual flux density of~${\mathbf{B}}^\mathrm{r}={B}^\mathrm{r}\mathbf{e}_3$ (${B}^\mathrm{r}=63\,$mT, see Methods and Supplementary Note S1), and is responsive to an external magnetic field, with a significant modification of the buckling pressure (Fig.~\ref{fig:Fig1}\textbf{d}) compared to the no-field case. When the field vector is parallel to the axis of magnetization (\textit{i.e.}, $\mathbf{B}^\mathrm{a}$ and $\mathbf{B}^\mathrm{r}$ are in the same direction), we observe an increase of the critical loads by $12\%$ and $24\%$ under the flux densities ${B}^\mathrm{a}=33\,$mT and ${B}^\mathrm{a}=66\,$mT, respectively. Meanwhile, the accompanying pressure drop at the onset of buckling is gradually reduced, eventually disappearing for~${B}^\mathrm{a}=66\,$mT. Therefore, the shell is strengthened by the applied field, and the buckling event becomes less catastrophic. By contrast, for the opposite field polarity (\textit{i.e.}, $\mathbf{B}^\mathrm{a}$ and $\mathbf{B}^\mathrm{r}$ are in opposite directions), the shell is weakened; the critical load decreases with a consequent more abrupt pressure drop past the buckling event (${B}^\mathrm{a}=\{-33, -66\}\,$mT in Fig.~\ref{fig:Fig1}\textbf{d}). These findings demonstrate that the intrinsic buckling strength of a magnetically active shell can be modified (increased or decreased), under an external magnetic field, on-demand. 

\begin{figure}[h!]
\centering
\includegraphics[width=\textwidth]{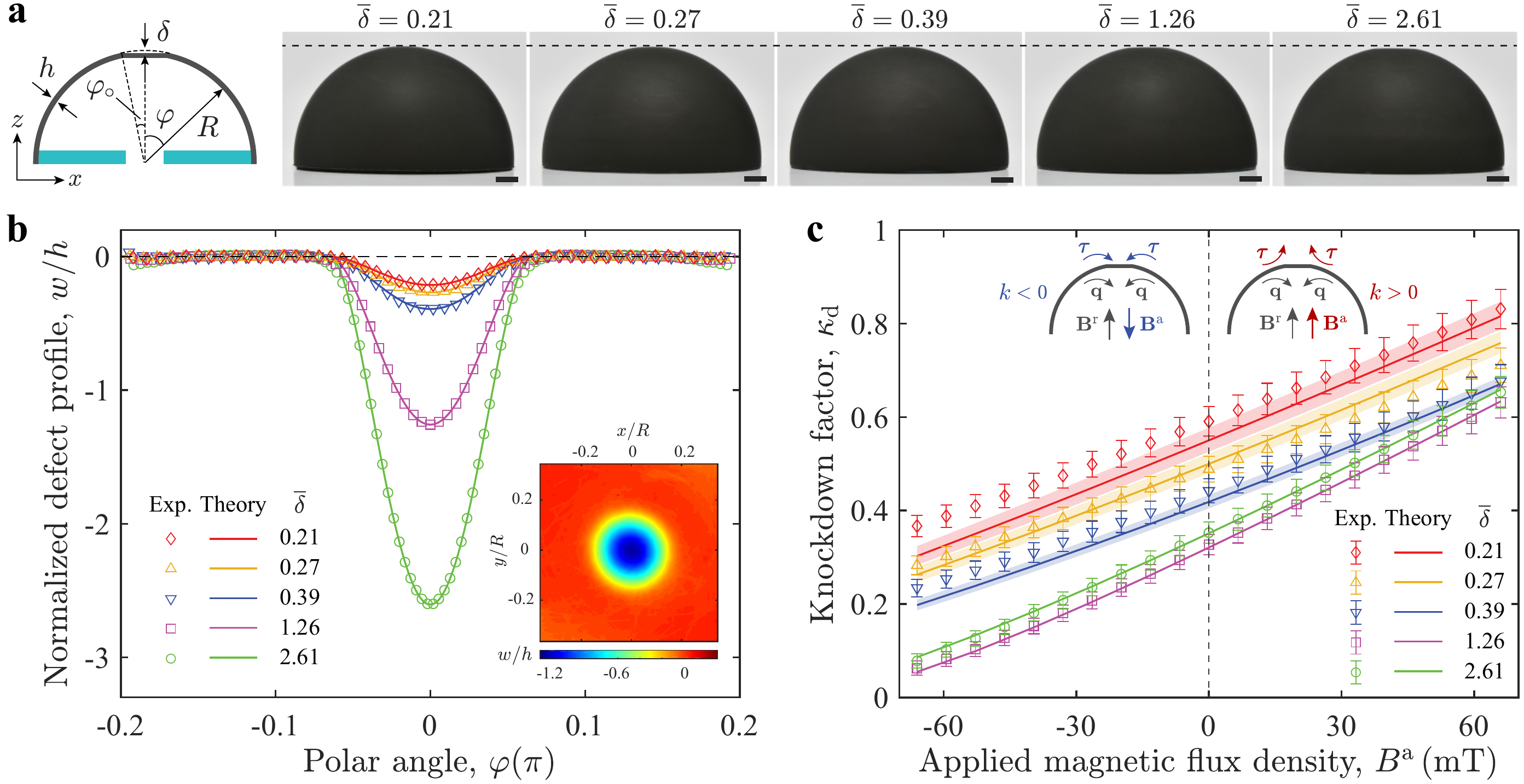}
\caption{\textbf{Tuning the knockdown factor of magnetic shells with different defect geometries.} \textbf{a}, Photographs of a series of magnetic shells with different defect amplitudes, $\overline{\delta}=\{0.21, 0.27, 0.39, 1.26, 2.61\}$, but same defect width, $\overline{\varphi}_\circ=3.2$. The schematic illustrates the geometric parameters of the shell. \textbf{b}, Axisymmetric 2D defect profiles normalized by shell thickness, $w/h$, for the shells presented in \textbf{a}. Symbols are the latitude-wise average of the corresponding 3D profiles measured through profilometry, and lines are profiles described by~${w}/{h}=-\overline{\delta}\left(1-{\varphi^2}/{\varphi_\circ^2}\right)^2\,(0\le\varphi\le\varphi_\circ)$ (see Supplementary Note S1 for details). Inset: Measured 3D defect profile of the shell with $\overline{\delta}=1.26$ and $\overline{\varphi}_\circ=3.2$. \textbf{c}, Knockdown factor, $\kappa_{\mathrm{d}}$, versus flux density of the applied field, ${B}^\mathrm{a}$, for shells containing the defects shown in \textbf{b}. Symbols and lines represent experimental results and theoretical predictions from the axisymmetric shell model, respectively. The error bars in the experimental data correspond to the accuracy of pressure measurements and the standard deviation of the measurements of shell thickness and Young's modulus; the error bands of theoretical predictions correspond to the standard deviation of the measurements of defect amplitude and shell thickness. Insets: Schematics of directions of the shell rotation vector, ${\mathbf{q}}$, and the magnetic torque vector, $\boldsymbol{\tau}=-{k}{\mathbf{q}}$, near the pole, for the cases of ${\mathbf{B}}^\mathrm{a}$ opposite (left) or parallel (right) to ${\mathbf{B}}^\mathrm{r}$.}
\label{fig:Fig2}
\end{figure}

To further explore the effect of magnetism on the buckling strength of pressurized spherical shells, we test shells with different defect geometries, over a range of external flux densities. As shown in the photographs of the specimens in Fig.~\ref{fig:Fig2}\textbf{a}, the defect amplitudes are varied as~$\overline{\delta}=\{0.21, 0.27, 0.39, 1.26, 2.61\}$ during manufacturing (Methods and Supplementary Note S1). The axisymmetric defect profiles, $w$, are presented in Fig.~\ref{fig:Fig2}\textbf{b}, defined as the radial deviation of the measured shell profile from a perfect hemisphere. Figure \ref{fig:Fig2}\textbf{c} presents the corresponding knockdown factor measurements, $\kappa_\mathrm{d}=\overline{p}_{\mathrm{max}}$, as a function of the external flux density, ${B}^\mathrm{a}$. Naturally, due to  imperfection sensitivity, the shells exhibit distinct knockdown factors for different values of~$\overline{\delta}$. However, in the presence of the magnetic field, we consistently observe an increasing or decreasing knockdown factor over the explored range of $\overline{\delta}$. Within the range of ${B}^\mathrm{a}$ accessible in our experiments, $\kappa_\mathrm{d}$ can be changed up to approximately $\pm30\%$, with respect to the no-field case. These experimental results demonstrate that the knockdown factor of a shell can be dynamically tuned, as an extrinsic quantity, by adjusting the polarity and strength of the applied magnetic field, via a robust mechanism that is insensitive to imperfections.

\noindent \textbf{Theory of axisymmetric hard-magnetic shells}\\
To rationalize the experimental results presented above, we develop a theoretical model that predicts the response of hard-magnetic axisymmetric thin shells under a combination of mechanical and magnetic loading. We consider the Helmholtz free energy of ideal hard-magnetic soft materials~\cite{Kim_Nature2018,Zhao_JMPS2019}, comprising an elastic energy term (related to material deformation) and a magnetic energy term (describing the work to align the residual magnetic vector along the external magnetic field). The shell model is developed by reducing this three-dimensional energy to the 1D profile curve of the middle surface of the shell (detailed derivation provided in Supplementary Note S3). The dimensional reduction of the Kirchhoff-Saint Venant elastic energy~\cite{Gurtin_MechanicsThermodynamicsContinua2010}, valid for small strains and large displacements, with the potential of live pressure was reported recently~\cite{Pezzulla_JAM2019}. In the present study, we focus on the magnetic energy term, which, for 3D scale-free materials, is written as~\cite{Bertotti_HysteresisMagnetism1998,Kim_Nature2018,Zhao_JMPS2019} $\mathcal{U}_\mathrm{m}=-{\mu_0}^{-1}\int\mathbf{F}{\mathbf{B}}^\mathrm{r}\cdot\mathbf{B}^\mathrm{a}\,\mathrm{d}V$, where $\mathbf{F}$ is the deformation gradient~\cite{Gurtin_MechanicsThermodynamicsContinua2010}. We describe the axisymmetric shell profile by the coordinates $(\mathring{\rho},\mathring{z})$, $\mathring{\rho}$ being the radial coordinate in the plane $(x,y)$ perpendicular to the axis of axisymmetry ($z$). Accented quantities (\textit{e.g.}, $\mathring{a}$) refer to the undeformed configuration of the shell. The reduced $1$D magnetic energy, normalized by~$\pi EhR^2/(4(1-\nu^2))$, can be derived as (Supplementary Note S3)
\begin{equation}
    \mathcal{\overline{U}}_\mathrm{m}=-\frac{8(1-\nu^2)}{R^2}\frac{{B}^\mathrm{r}B^\mathrm{a}}{\mu_0 E}\int_0^{\pi/2}\mathcal{F}(\mathring{\rho},_\varphi, \mathring{z},_\varphi, \rho,_\varphi, z,_\varphi)\ \mathring{a}\mathrm{d}\varphi\,,
    \label{eq:magnetic_energy_1D} 
\end{equation}
where $\mathring{a}$ is the area measure, and~$\mathcal{F}$ is a dimensionless function that depends on the initial and deformed configurations of the shell. From Eq.~(\ref{eq:magnetic_energy_1D}), we identify a magneto-elastic parameter
\begin{equation}
    \lambda_\mathrm{m}=\frac{{B}^\mathrm{r}B^\mathrm{a}}{\mu_0 E}\,,
    \label{eq:lambda_m} 
\end{equation}
which represents the intrinsic magneto-elastic coupling of the system. Equilibrium equations can be generated by minimizing the total energy for all the possible displacements of the shell, which were solved via the Newton-Raphson method (see Methods and Supplementary Note S3.1.4)~\cite{Pezzulla_JAM2019}.

It is important to highlight that, in our simulations of the 1D model presented above, we use the geometric and physical parameters of the system measured in experiments (see Methods), without any fitting parameters. The defect profile in the region $0\le\varphi\le\varphi_\circ$ is described analytically by ${w}/{h}=-\overline{\delta}\left(1-{\varphi^2}/{\varphi_\circ^2}\right)^2$ (Fig.~\ref{fig:Fig2}\textbf{b}), derived based on a simple plate model (Supplementary Note S1). Excellent agreement is found between theory and experiments (Fig.~\ref{fig:Fig2}\textbf{c}). The variation of the knockdown factor for shells with different defect geometries under the magnetic field is accurately predicted by our shell model, which is, therefore, able to describe the intricate coupling between elasticity, magnetism, and the nonlinear mechanics of thin shells.

\noindent \textbf{Physical interpretation of the reduced magnetic energy}\\
Even though our theoretical model can predict the buckling strength of shells, the reduced magnetic energy is highly nonlinear, and the mechanism underlying the change of knockdown factors still needs to be clarified. Therefore, we proceed to expand the integrand of the reduced magnetic energy density in Eq.~\eqref{eq:magnetic_energy_1D}, $\mathcal{F}(\mathring{\rho},_\varphi, \mathring{z},_\varphi, \rho,_\varphi, z,_\varphi)$, up to second order in the displacement field (Supplementary Note S3). By examining the role of each term in the buckling instability, we conclude that only the following second-order term dominates the buckling of the shell,
\begin{equation}
    \mathcal{\overline{U}}_\mathrm{m}^{(2)\mathrm{A}}=-\frac{8(1-\nu^2)}{R^2}\int_0^{\pi/2}\frac{1}{2} \boldsymbol{\tau}\cdot\mathbf{q}\ \mathring{a} \mathrm{d}\varphi\,,
\end{equation}
where $\mathbf{q}$ is the rotation vector of a material fiber of the shell~\cite{Niordson_ShellTheory1985,Pezzulla_PRL2018}, and  $\boldsymbol{\tau}=-{k}(\varphi)\mathbf{q}$ can be interpreted as a distributed (dimensionless) torque applied by the external magnetic field. This torque is a linear function of the shell rotation with a deformation-independent pre-factor ${k}=\lambda_\mathrm{m}\mathring{\chi}(\varphi)$, which we can interpret as the (dimensionless) stiffness of distributed rotational springs, where $\mathring{\chi}(\varphi)={\mathring{\rho},_\varphi^2}/{(\mathring{\rho},_\varphi^2+\mathring{z},_\varphi^2)}$. 

Under pressure loading, material fibers tend to rotate, thereby increasing the magnetic torque (Supplementary Note S3). Whether the torque reacts to restore the undeformed orientation of the material fibers or to rotate them further away from the initial orientation, depends on the sign of the equivalent stiffness $k$ (insects in Fig.~\ref{fig:Fig2}\textbf{c}). When $\mathbf{B}^{\mathrm{a}}$ and ${\mathbf{B}}^{\mathrm{r}}$ are in the same direction, the stiffness ${k}$ is positive, thus ensuring that $\boldsymbol{\tau}$ is opposite to $\mathbf{q}$, thereby counteracting buckling. As a result, we observe the strengthening of shells with increasing critical loads (cases with ${B}^\mathrm{a}>0$ in Fig.~\ref{fig:Fig2}\textbf{c}). Conversely, when $\mathbf{B}^{\mathrm{a}}$ and ${\mathbf{B}}^{\mathrm{r}}$ and in opposite directions, the negative stiffness (${k}<0$) leads to a torque $\boldsymbol{\tau}$ that acts to increase the rotation $\mathbf{q}$. Indeed, in this regime, buckling occurs at lower pressure levels (cases with ${B}^\mathrm{a}<0$ in Fig.~\ref{fig:Fig2}\textbf{c}).

\noindent \textbf{Scaling analysis of the change of knockdown factor}\\
With a physical interpretation of the magnetic energy for shells at hand, we now employ scaling arguments to more clearly rationalize how the knockdown factor is modified by the magnetic field for different radius-to-thickness ratios.
Since the magnetic energy interacts with the live pressure potential to alter the knockdown factor, we balance the two energy terms. The dimensionless magnetic energy can be shown to scale as~$\overline{\mathcal{U}}_\mathrm{m}\sim\lambda_\mathrm{m}{h}/{R}$, while the change of knockdown factor with respect to the non-magnetic case, $\Delta\kappa_\textup{d}=\kappa_\mathrm{d}-\kappa_{\mathrm{d}}|_{{B}^\mathrm{a}=0}$, scales as~$\Delta \kappa_\textup{d} \sim \Delta p/E(R/h)^2$ (Supplementary Note S3). By balancing the scalings for the potential of~$\Delta p$, that is~$\overline{\mathcal{U}}_{\Delta p}=\Delta p/E$, and the magnetic energy~$\overline{\mathcal{U}}_\mathrm{m}$, we find $\Delta p/E\sim\lambda_\textup{m}{h}/{R}$, translating into
\begin{equation}
    \Delta\kappa_\textup{d}\sim \frac{\Delta p}{E}\left(\frac{R}{h}\right)^2\sim\lambda_\textup{m}\frac{R}{h}\,.
    \label{eq:scaling_law_buckling}
\end{equation}
From Eq.~(\ref{eq:scaling_law_buckling}), we define the \emph{magneto-elastic buckling number},
\begin{equation}
    \Lambda_\mathrm{m}=\lambda_\textup{m} \frac{R}{h},
    \label{eq:magnetoelasticbucklingparameter}
\end{equation}
which governs the knockdown factor under combined pressure and magnetic loading of magnetic shells. The scaling $\Delta\kappa_\textup{d} \sim \Lambda_\mathrm{m}$ provides a scale-invariant description of how the magnetic field modifies the buckling pressure for shells with different radius-to-thickness ratios. 

\begin{figure}[t]
\centering
\includegraphics[width=0.8\textwidth]{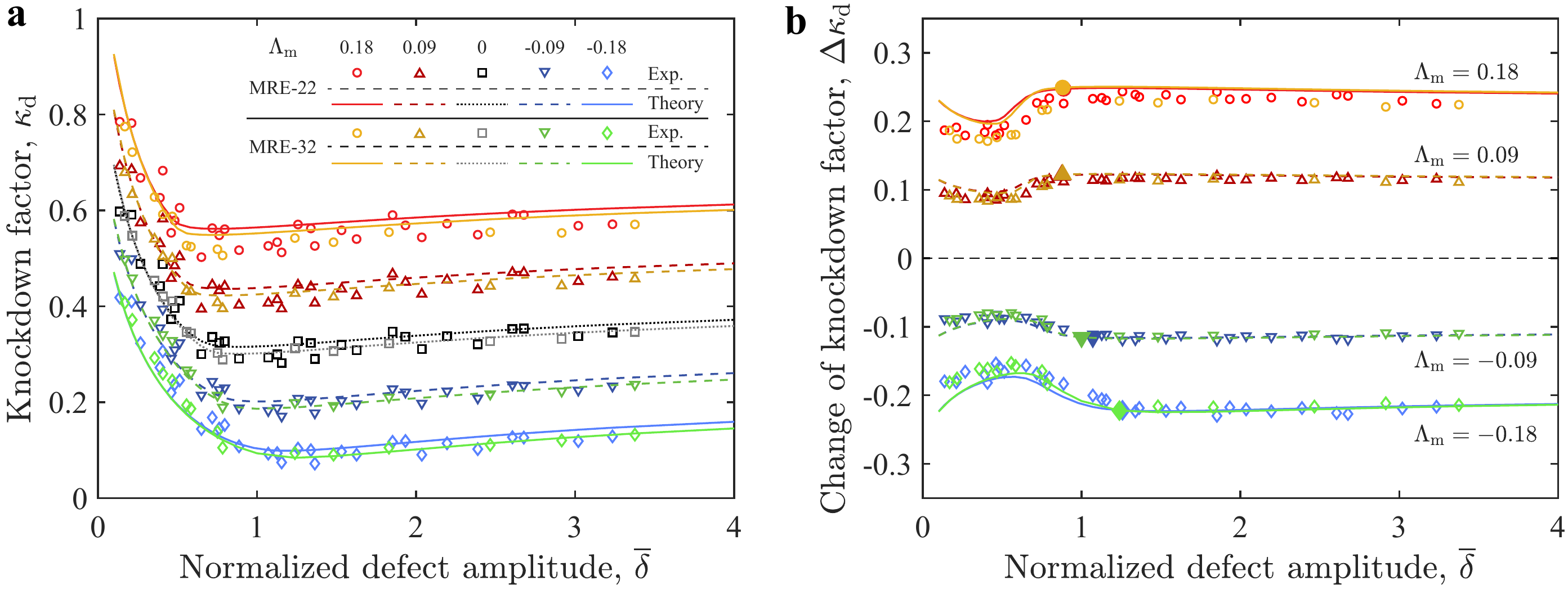}
\caption{\textbf{Imperfection sensitivity of the pressure buckling of magnetic shells.} \textbf{a}, Knockdown factor, $\kappa_\mathrm{d}$, and, \textbf{b}, its change under the magnetic field, $\Delta\kappa_\mathrm{d}=\kappa_\mathrm{d}-\kappa_{\mathrm{d}}|_{{B}^\mathrm{a}=0}$ are plotted versus normalized defect amplitude $\overline{\delta}$. The tested shell specimens are made of MRE-22 or MRE-32  ($E=1.15\,$MPa, $R/h=79.1$ for MRE-22 and $E=1.69\,$MPa, $R/h=91.3$ for MRE-32; see Methods). The magneto-elastic buckling number $\Lambda_\mathrm{m}=\{0.18, 0.09, 0, -0.09, -0.18\}$ is varied by adjusting the external flux density ${B}^\mathrm{a}=\{-53, -26, 0, 26, 53\}\,$mT for the MRE-22 shells and ${B}^\mathrm{a}=\{-66, -33, 0, 33, 66\}\,$mT for the MRE-32 shells. Experimental results are represented by open symbols, and lines are predictions by the axisymmetric shell model. The closed symbols on each line in \textbf{b} represent the onset of the plateau of $\Delta\kappa_\mathrm{d}$, determined as the first point when the derivative of $\Delta\kappa_\mathrm{d}$ with respect to $\overline{\delta}$ is smaller than 1\%.}
\label{fig:Fig3}
\end{figure}

\noindent \textbf{Robustness of the mechanism to geometric imperfections}\\
%\\
We set out to investigate the role of imperfections in the buckling of magnetic shells, which we now address with a more systematic parametric exploration. We fabricate shells over a wide range of the defect amplitude ($0.1\le\overline{\delta}\le4$) and measure the corresponding knockdown factors, $\kappa_\mathrm{d}$, from buckling tests. In parallel, we run 1D simulations for the same material and geometrical properties. Figure~\ref{fig:Fig3}\textbf{a} illustrates the relationship between $\kappa_\mathrm{d}$ and $\overline{\delta}$ at different levels of external flux density, included in the magneto-elastic buckling number, $\Lambda_\mathrm{m}$, of Eq.~\eqref{eq:magnetoelasticbucklingparameter}. The signature of imperfection sensitivity is still observed in the presence of the magnetic field: $\kappa_\mathrm{d}$ decreases dramatically when the defect amplitude increases in the regime of relatively small defects ($0<\overline{\delta}<1$). Surprisingly, the results of the change in knockdown factor under the applied magnetic field ($\Delta\kappa_\mathrm{d}$) presented in Fig.~\ref{fig:Fig3}\textbf{b} are significantly less sensitive to defects; $\Delta\kappa_\mathrm{d}$ becomes approximately constant for $\overline{\delta}>1$. This finding suggests that the magnetic interaction between the shell and the external field is nearly unaffected by the intrinsic imperfection sensitivity, ensuring a robust tuning of the knockdown factor. 

Moreover, the results in Fig.~\ref{fig:Fig3}\textbf{a},\textbf{b} include data for two sets of shells with a different combination of material and geometric properties ($E=1.15\,$MPa, $R/h=79.1$ for MRE-22 shells and $E=1.69\,$MPa, $R/h=91.3$ for MRE-32 shells; see Methods). Still, the $\kappa_\mathrm{d}$ and $\Delta\kappa_\mathrm{d}$ data collapse for these two sets, given that the value of $\Lambda_\mathrm{m}$ is the same for both. This collapse supports  $\Lambda_\mathrm{m}$ as the governing parameter, combining the mechanical, magnetic, and geometrical properties of the system. Throughout the analysis, the theory is in excellent agreement with experiments.
%, suggesting the model as a fast and effective tool for the design of magnetic shells.

\begin{figure}[b!]
\centering
\includegraphics[width=0.8\textwidth]{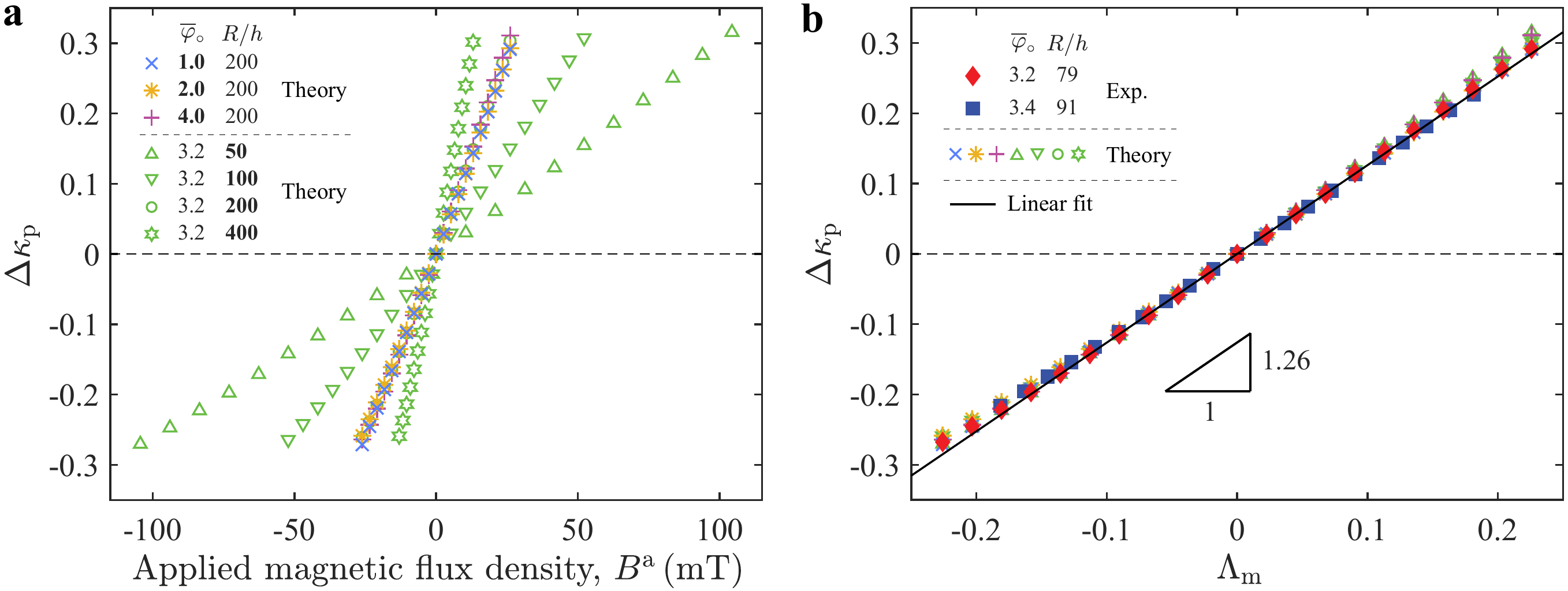}
\caption{\textbf{Plateau of the change of knockdown factor.} \textbf{a}, Plateau change of knockdown factor, $\Delta\kappa_\mathrm{p}$, as a function of external magnetic flux density, ${B}^\mathrm{a}$, is presented for shells with different defect widths, $\overline{\varphi}_\circ=\{1.0, 2.0, 3.2, 4.0\}$, and in different radius-to-thickness ratios, $R/h=\{50, 100, 200, 400\}$ (as detailed in the legend). Results are obtained by the axisymmetric shell model. \textbf{b}, Collapse of the data by plotting $\Delta\kappa_\mathrm{p}$ versus the magneto-elastic buckling number $\Lambda_\mathrm{m}$. Experimental and theoretical results are juxtaposed, showing excellent agreement. A linear fit on the data indicates a slope of $1.26 \pm 0.01$.}
\label{fig:Fig4}
\end{figure}

A robust way to quantify the effect of the magnetic field on the change in knockdown factor is to focus on the plateau in Fig.~\ref{fig:Fig3}\textbf{b}, $\Delta\kappa_\mathrm{p}$, defined as the average of $\Delta\kappa_\mathrm{d}$ over the extent of the plateau-like region. Figure~\ref{fig:Fig4}\textbf{a} (open symbols) shows predictions from our shell model for $\Delta\kappa_\mathrm{p}$ versus the external flux density, ${B}^\mathrm{a}$, for shells with different radius-to-thickness ratios and defect widths.
While the radius-to-thickness ratio strongly influences the change in knockdown factor, the defect width has little effect on it (Supplementary Note 4). In Fig.~\ref{fig:Fig4}\textbf{b}, we plot the raw data of Fig.~\ref{fig:Fig4}\textbf{a} as a function of the magneto-elastic buckling number $\Lambda_\mathrm{m}$, finding that the different results for shells with different geometries fall onto a master curve. This collapse demonstrates that the plateau change of knockdown factor is governed by the single parameter $\Lambda_\mathrm{m}$, independently of the geometry of the defect. The superposition of experimental results on the theoretical predictions further validates this collapse. Moreover, consistently with the scaling in Eq.~\eqref{eq:scaling_law_buckling}, a linear relationship is found between $\Delta\kappa_\mathrm{p}$ and $\Lambda_\mathrm{m}$, with a slope of $1.26 \pm 0.01$ obtained via linear fitting. This master curve serves as a concrete design guideline for magnetic shells,  summarizing the effect of the magnetic field in tuning the buckling strength of pressurized spherical shells with different material and geometrical properties.

In closing, we have shown that the buckling strength of shells can be dynamically tuned by exploiting the interplay between mechanics and magnetism. The proposed mechanism represents a robust way to gain control on a property of thin shells that has long been regarded as intrinsic, the knockdown factor. By performing precision experiments on hard-magnetic elastomeric shells and developing a theoretical model, we unveiled the essential feature at the base of the mechanism, a distributed torque induced by the magnetic field throughout the shell. Moreover, we showed that a dimensionless quantity, the magneto-elastic buckling number, emerges as the key governing parameter, summarizing the geometrical, mechanical, and magnetic properties of the system. We envision that our mechanism can also be used to gain control on the structural life of a shell, by tuning the magnetic field as critical conditions approach. More generally, we believe that the approach of coupling mechanical deformation and magnetic fields, already successfully applied for beams~\cite{Lum_PNAS2016,Zhao_JMPS2019,Kim_SciRobot2019,Wang_JMPS2020}, films~\cite{Psarra_JMPS2019}, and soft materials~\cite{Kim_Nature2018}, and employed here to tune the knockdown factor of shells, can be extended to modify the instabilities of other structures such as rods, plates, and non-axisymmetric shells~\cite{Loukaides_IntJSmartNanoMater2014,Seffen_SmartMaterStruct2016}, for which research efforts are currently ongoing. We anticipate that magnetic structures with predictable and tunable buckling behavior will enrich the design of advanced materials with new functionalities.

\section*{Methods}

\noindent \textbf{Preparation of the MRE material.} Our experimental samples were fabricated using a MRE material composed of a mixture of hard-magnetic NdPrFeB particles (average size of $5\,\mu$m, MQFP-15-7-20065-089, Magnequench) and Vinylpolysiloxane  (VPS, Elite Double, Zhermack), a silicone-based polymer. MREs made of VPS Elite Double 22 or VPS Elite Double 32 are referenced throughout the text as MRE-22 or MRE-32, respectively. We prepared the MRE with the following steps. First, the initially non-magnetized NdPrFeB particles were mixed with the liquid VPS base (1:1 mass ratio) using a centrifugal mixer (ARE-250, Thinky Corporation); for $40\,$s at $2000\,$rpm (clockwise), and another $20\,$s at $2200\,$rpm (counterclockwise). Secondly, the solution was degassed in a vacuum chamber (absolute pressure below $8\,$mbar), and then cooled down to room temperature ($22\pm0.4\,^\circ$C) to avoid any changes of viscosity due to thermal disturbances from the previous steps. Thirdly, VPS catalyst was added to the mixture, with a ratio of 1:1 in weight to the VPS base. After another mixing step for $20\,$s at $2000\,$rpm (clockwise) followed by $10\,$s at $2200\,$rpm (counterclockwise), the final mixture was ready for sample fabrication. The final mass concentrations of NdPrFeB particles and VPS polymer were $33.3\,\mathrm{wt}\%$ and $66.7\,\mathrm{wt}\%$, respectively. Pouring of the liquid MRE mixture onto the mold during shell fabrication (see below) was done after a set waiting time ($100\,$s for MRE-22 and $20\,$s for MRE-32), so as to increase the viscosity to the desired value. After the above preparation steps, the curing of the polymer mixture occurred in approximately 20 min at room temperature.

\noindent \textbf{Physical properties of the MRE.} The densities of MRE-22 and MRE-32 were measured using a pycnometer to be $\rho_\mathrm{MRE-22}=1.61\pm0.05\,\mathrm{g/cm^3}$ and $\rho_\mathrm{MRE-32}=1.63\pm0.03\,\mathrm{g/cm^3}$, respectively. The density of the NdPrFeB particles is $7.61\,\mathrm{g/cm^3}$ (provided by the supplier). Under the concentrations of the prepared mixture, the volume fraction of NdPrFeB particles was calculated to be 7.05\% for MRE-22 and 7.14\% for MRE-32. When saturated, the effective residual flux density of our MRE, ${B}^\mathrm{r}$, was assumed to be the volume-average of the total residual flux density of individual NdPrFeB particles ($0.90\,$T, as reported by the supplier). Given the chosen volume fraction, ${B}^\mathrm{r} = 63.2\,$mT for MRE-22 and ${B}^\mathrm{r} = 64.0\,$mT for MRE-32. The Young's moduli of MRE-22 and MRE-32 were measured to be $E=1.15\pm0.04\,\mathrm{MPa}$, and $E=1.69\pm0.06\,\mathrm{MPa}$, respectively, using a combination of cantilever and tensile tests. 

\noindent \textbf{Fabrication of the magneto-active shell specimens.} Our shells were fabricated by coating the underside of a flexible negative spherical mold (radius $25.4\,$mm) with liquid MRE; see the schematic in Supplementary Fig.~S1 and Supplementary Note S1 for details. Whereas our technique was inspired by previous work~\cite{Lee_NatCommun2016,Lee_JAM2016}, our molds also contained a soft spot (thin circular region) at the pole to produce a precisely-engineered defect. With the negative spherical surface of the mold facing down, the gravity-driven viscous flow of the polymer yielded a thin layer of MRE on the mold. Before the MRE fully cured ($\approx13\,$min after pouring), the mold was depressurized from within using a syringe pump. As a result, the soft spot of the mold deformed to produce an axisymmetric geometric imperfection at the pole of the shell. This defect becomes `frozen' in the shell upon curing. The amplitude of the defect can be systematically varied by adjusting the pressure imposed on the mold during curing (Supplementary Fig.~S2). The width of the defect can be varied by changing the pillar used in mold manufacture with a different radius (Supplementary Note S1). The amplitude and width of the defect can be set independently. To make the shell magnetically active, we saturated the NdPrFeB particles, already in the solidified MRE, using an applied uniaxial magnetic field ($4.4\,$T, perpendicular to the equatorial plane) generated by an impulse magnetizer (IM-K-010020-A, Magnet-Physik Dr. Steingroever GmbH). 

\noindent \textbf{Geometry of the shells.} The radius of the  shells ($R=25.4\,$mm) is set by the mold used during fabrication. Their thickness was characterized using a microscope (VHX-5000, Keyence Corporation) after cutting off narrow strips ($\approx2\times6\,\mathrm{mm}^2$) near the pole. The average measured thickness was $h=321.2\pm5.1\,\mathrm{\mu m}$ and $h=278.2\pm2.8\,\mathrm{\mu m}$ for the MRE-22 and the MRE-32 shells, respectively. Hence, the corresponding radius-to-thickness ratios were $R/h=79.1$ and $R/h=91.3$. The 3D profile of the outer surface of each fabricated imperfect shell was characterized using an optical profilometer (VR-3200, Keyence Corporation). The geometry of the defect was computed as the radial distance between the measured profile and the corresponding best-fit spherical surface. Specifically, the amplitude, $\delta$, and half angular width, $\varphi_\circ$, of the defect were determined by fitting the analytical description, ${w}/{h}=-\overline{\delta}\left(1-{\varphi^2}/{\varphi_\circ^2}\right)^2\,(0\le\varphi\le\varphi_\circ)$, to the experimental measurements (additional details provided in Supplementary Note S1). The ranges of geometric parameters for our specimens are as follows: $0.14 \leq \overline{\delta}\leq 3.2$ and $\varphi_\circ=11.7\pm0.1\,^\circ$ ($\overline{\varphi}_\circ=3.2$) for the MRE-22 shells; and $0.17 \le \overline{\delta} \le 3.4$ and $\varphi_\circ=11.6\pm0.1\,^\circ$ ($\overline{\varphi}_\circ=3.4$) for the MRE-32 shells.

\noindent \textbf{Generation of the magnetic field.}  A uniaxial uniform magnetic field was generated in the central region of two identical customized multi-turn circular coils (square cross-section, inner diameter $72\,$mm, outer diameter $114\,$mm, and height $21\,$mm), configured in the Helmholtz configuration. The coils were set concentrically, with a centre-to-centre axial distance of $46.5\,$mm, and connected in series, powered by a DC power supply providing a maximum current/power of $25\,$A/$1.5\,$kW (EA-PSI 9200-25 T, EA-Elektro-Automatik GmbH). The flux density of the magnetic field was varied in the range $-66\,\mathrm{mT} \le {B}^\mathrm{a} \le 66\,\mathrm{mT} $ by adjusting the current output of the power supply, from $-10\,$A to $10\,$A. The flux density was measured using a Teslameter (FH 55, Magnet-Physik Dr. Steingroever GmbH), along both the axial and radial directions. In parallel to the experiments, we simulated the generated field using the \textit{Magnetic Fields} interface embedded in the commercial package COMSOL Multiphysics (v. 5.2, COMSOL Inc.), based on Amp\`{e}re's Law (see Supplementary Note S2 for details). Excellent agreement was found between experiments and simulations (Supplementary Fig.~S3).

\noindent \textbf{Buckling tests.} To quantify the critical buckling conditions of our magnetic shells, we positioned each shell in between the Helmholtz coils under a steady magnetic field, and then depressurized it under prescribed volume conditions by a pneumatic loading system (see Supplementary Note S2 for details). The applied pressure was monitored by a pressure sensor (785-HSCDRRN002NDAA5, Honeywell International Inc.). The level of depressurization was increased, under a steady magnetic field, until the shell buckled. The critical buckling pressure was defined as the maximum value of the loading curve (pressure \textit{vs.} imposed volume change).

\noindent \textbf{Theory and numerics.} The reduced 1D energy~$\overline{\mathcal{U}}$, representing the sum of the elastic and magnetic energies of the system, together with the potential of the live pressure, is obtained by means of dimensional reduction (see Supplementary Note S3 for details). Equilibrium equations are generated by minimizing the total energy for all possible displacements of the shell, and solved via the Newton-Raphson method using the commercial package COMSOL Multiphysics (v. 5.2, COMSOL Inc.). Details of the implementation of our numerical procedure are provided in Ref.~\cite{Pezzulla_JAM2019}.

\section*{Acknowledgements} 

We thank Selman Sakar and Lucio Pancaldi-Giubbini for providing the coils. We acknowledge Benjamin Rahm for help with preliminary experiments. A.A. is grateful to the support from the Federal Commission for Scholarships for Foreign Students (FCS) through Swiss Government Excellence Scholarship (Grant No. 2019.0619).

\noindent This is a preprint of an article published in \textit{Nature Communications}. The final authenticated version is available online at: \url{https://doi.org/10.1038/s41467-021-22776-y}.

\section*{Author contributions}
D.Y. and P.M.R. conceived the project. D.Y., L.C. and A.A. performed experiments and analyzed data. M.P. developed the theoretical model and performed the scaling analysis. M.P. and D.Y. performed simulations and analyzed data. P.M.R. supervised the research. D.Y., M.P., A.A. and P.M.R. wrote the paper.

\section*{Competing interests}
The authors declare no competing interests.

\section*{Additional information}
%Supplementary information is available for this paper at.\\
Correspondence and requests for materials should be addressed to P.M.R.

\bibliographystyle{naturemag.bst} 
\bibliography{references.bib}

\begin{thebibliography}{10}
\expandafter\ifx\csname url\endcsname\relax
  \def\url#1{\texttt{#1}}\fi
\expandafter\ifx\csname urlprefix\endcsname\relax\def\urlprefix{URL }\fi
\providecommand{\bibinfo}[2]{#2}
\providecommand{\eprint}[2][]{\url{#2}}

\bibitem{Koiter_NonlinearBuckling1969}
\bibinfo{author}{Koiter, W.~T.}
\newblock \bibinfo{title}{The nonlinear buckling behavior of a complete
  spherical shell under uniform external pressure, parts i, ii, iii \& iv}.
\newblock \emph{\bibinfo{journal}{Proc. Kon. Ned. Ak. Wet.}}
  \textbf{\bibinfo{volume}{B72}}, \bibinfo{pages}{40--123}
  (\bibinfo{year}{1969}).

\bibitem{Niordson_ShellTheory1985}
\bibinfo{author}{Niordson, F.~I.}
\newblock \emph{\bibinfo{title}{Shell Theory}}.
\newblock North-Holland Series in Applied Mathematics and Mechanics
  (\bibinfo{publisher}{Elsevier Science}, \bibinfo{year}{1985}).

\bibitem{Hutchinson_EML2020}
\bibinfo{author}{Hutchinson, J.~W.}
\newblock \bibinfo{title}{{EML} {Webinar} overview: {New} developments in shell
  stability}.
\newblock \emph{\bibinfo{journal}{Extreme Mech. Lett.}}
  \textbf{\bibinfo{volume}{39}}, \bibinfo{pages}{100805}
  (\bibinfo{year}{2020}).

\bibitem{Hilburger_AIAA2012}
\bibinfo{author}{Hilburger, M.~W.}
\newblock \bibinfo{title}{Developing the next generation shell buckling design
  factors and technologies}.
\newblock In \emph{\bibinfo{booktitle}{53rd {AIAA}/{ASME}/{ASCE}/{AHS}/{ASC}
  {Structures}, {Structural} {Dynamics} and {Materials} {Conference}}},
  Structures, {Structural} {Dynamics}, and {Materials} and {Co}-located
  {Conferences} (\bibinfo{publisher}{American Institute of Aeronautics and
  Astronautics}, \bibinfo{address}{Honolulu, HI}, \bibinfo{year}{2012}).

\bibitem{Lidmar_PRE2003}
\bibinfo{author}{Lidmar, J.}, \bibinfo{author}{Mirny, L.} \&
  \bibinfo{author}{Nelson, D.~R.}
\newblock \bibinfo{title}{Virus shapes and buckling transitions in spherical
  shells}.
\newblock \emph{\bibinfo{journal}{Phys. Rev. E}} \textbf{\bibinfo{volume}{68}},
  \bibinfo{pages}{051910} (\bibinfo{year}{2003}).

\bibitem{sacanna2010}
\bibinfo{author}{Sacanna, S.}, \bibinfo{author}{Irvine, W.},
  \bibinfo{author}{Chaikin, P.} \& \bibinfo{author}{Pine, D.}
\newblock \bibinfo{title}{Lock and key colloids}.
\newblock \emph{\bibinfo{journal}{Nature}} \textbf{\bibinfo{volume}{464}},
  \bibinfo{pages}{575--578} (\bibinfo{year}{2010}).

\bibitem{Datta_PRL2012}
\bibinfo{author}{Datta, S.~S.} \emph{et~al.}
\newblock \bibinfo{title}{Delayed buckling and guided folding of inhomogeneous
  capsules}.
\newblock \emph{\bibinfo{journal}{Phys. Rev. Lett.}}
  \textbf{\bibinfo{volume}{109}}, \bibinfo{pages}{134302}
  (\bibinfo{year}{2012}).

\bibitem{Vian_SoftMatter2019}
\bibinfo{author}{Vian, A.} \& \bibinfo{author}{Amstad, E.}
\newblock \bibinfo{title}{Mechano-responsive microcapsules with uniform thin
  shells}.
\newblock \emph{\bibinfo{journal}{Soft Matter}} \textbf{\bibinfo{volume}{15}},
  \bibinfo{pages}{1290--1296} (\bibinfo{year}{2019}).

\bibitem{Katifori_PNAS2010}
\bibinfo{author}{Katifori, E.}, \bibinfo{author}{Alben, S.},
  \bibinfo{author}{Cerda, E.}, \bibinfo{author}{Nelson, D.~R.} \&
  \bibinfo{author}{Dumais, J.}
\newblock \bibinfo{title}{Foldable structures and the natural design of pollen
  grains}.
\newblock \emph{\bibinfo{journal}{Proc. Natl. Acad. Sci. U.S.A.}}
  \textbf{\bibinfo{volume}{107}}, \bibinfo{pages}{7635--7639}
  (\bibinfo{year}{2010}).

\bibitem{Forterre_Nature2005}
\bibinfo{author}{Forterre, Y.}, \bibinfo{author}{Skotheim, J.~M.},
  \bibinfo{author}{Dumais, J.} \& \bibinfo{author}{Mahadevan, L.}
\newblock \bibinfo{title}{How the {Venus} flytrap snaps}.
\newblock \emph{\bibinfo{journal}{Nature}} \textbf{\bibinfo{volume}{433}},
  \bibinfo{pages}{421--425} (\bibinfo{year}{2005}).

\bibitem{misbah2016}
\bibinfo{author}{Misbah, C.}
\newblock \emph{\bibinfo{title}{Complex Dynamics and Morphogenesis}}
  (\bibinfo{publisher}{Springer}, \bibinfo{year}{2016}).

\bibitem{Tsien_AeronauticalSciences1942}
\bibinfo{author}{Tsien, H.-S.}
\newblock \bibinfo{title}{A theory for the buckling of thin shells}.
\newblock \emph{\bibinfo{journal}{Journal of the Aeronautical Sciences}}
  \textbf{\bibinfo{volume}{9}}, \bibinfo{pages}{373--384}
  (\bibinfo{year}{1942}).

\bibitem{Hutchinson_JAM1967}
\bibinfo{author}{Hutchinson, J.~W.}
\newblock \bibinfo{title}{Imperfection sensitivity of externally pressurized
  spherical shells}.
\newblock \emph{\bibinfo{journal}{J. Appl. Mech.}}
  \textbf{\bibinfo{volume}{34}}, \bibinfo{pages}{49--55}
  (\bibinfo{year}{1967}).

\bibitem{Carlson_ExpMech1967}
\bibinfo{author}{Carlson, R.~L.}, \bibinfo{author}{Sendelbeck, R.~L.} \&
  \bibinfo{author}{Hoff, N.~J.}
\newblock \bibinfo{title}{Experimental studies of the buckling of complete
  spherical shells}.
\newblock \emph{\bibinfo{journal}{Exp. Mech.}} \textbf{\bibinfo{volume}{7}},
  \bibinfo{pages}{281--288} (\bibinfo{year}{1967}).

\bibitem{Zoelly_PhDThesis1915}
\bibinfo{author}{Zoelly, R.}
\newblock \emph{\bibinfo{title}{Ueber ein knickungsproblem an der
  kugelschale}}.
\newblock \bibinfo{type}{Ph.{D}. thesis}, \bibinfo{school}{ETH Z\"urich},
  \bibinfo{address}{Z\"urich, Switzerland} (\bibinfo{year}{1915}).

\bibitem{Elishakoff2014}
\bibinfo{author}{Elishakoff, I.}
\newblock \emph{\bibinfo{title}{Resolution of the Twentieth Century Conundrum
  in Elastic Stability}} (\bibinfo{publisher}{World Scientific Publishing,
  Singapore}, \bibinfo{year}{2014}).

\bibitem{Lee_JAM2016}
\bibinfo{author}{Lee, A.}, \bibinfo{author}{L\'opez~Jim\'enez, F.},
  \bibinfo{author}{Marthelot, J.}, \bibinfo{author}{Hutchinson, J.~W.} \&
  \bibinfo{author}{Reis, P.~M.}
\newblock \bibinfo{title}{The geometric role of precisely engineered
  imperfections on the critical buckling load of spherical elastic shells}.
\newblock \emph{\bibinfo{journal}{J. Appl. Mech.}}
  \textbf{\bibinfo{volume}{83}}, \bibinfo{pages}{111005}
  (\bibinfo{year}{2016}).

\bibitem{Lee_NatCommun2016}
\bibinfo{author}{Lee, A.} \emph{et~al.}
\newblock \bibinfo{title}{Fabrication of slender elastic shells by the coating
  of curved surfaces}.
\newblock \emph{\bibinfo{journal}{Nat. Commun.}} \textbf{\bibinfo{volume}{7}},
  \bibinfo{pages}{11155} (\bibinfo{year}{2016}).

\bibitem{Hutchinson_ProcMathPhysEngSci2016}
\bibinfo{author}{Hutchinson, J.~W.}
\newblock \bibinfo{title}{Buckling of spherical shells revisited}.
\newblock \emph{\bibinfo{journal}{Proc. Math. Phys. Eng. Sci.}}
  \textbf{\bibinfo{volume}{472}}, \bibinfo{pages}{20160577}
  (\bibinfo{year}{2016}).

\bibitem{Hutchinson_PhilosTransRoyalSocA2017}
\bibinfo{author}{Hutchinson, J.~W.} \& \bibinfo{author}{Thompson, J. M.~T.}
\newblock \bibinfo{title}{Nonlinear buckling behaviour of spherical shells:
  barriers and symmetry-breaking dimples}.
\newblock \emph{\bibinfo{journal}{Philos. Trans. Royal Soc. A}}
  \textbf{\bibinfo{volume}{375}}, \bibinfo{pages}{20160154}
  (\bibinfo{year}{2017}).

\bibitem{Jimenez_JAM2017}
\bibinfo{author}{L\'opez~Jim\'enez, F.}, \bibinfo{author}{Marthelot, J.},
  \bibinfo{author}{Lee, A.}, \bibinfo{author}{Hutchinson, J.~W.} \&
  \bibinfo{author}{Reis, P.~M.}
\newblock \bibinfo{title}{Technical brief: {Knockdown} factor for the buckling
  of spherical shells containing large-amplitude geometric defects}.
\newblock \emph{\bibinfo{journal}{J. Appl. Mech}}
  \textbf{\bibinfo{volume}{84}}, \bibinfo{pages}{034501}
  (\bibinfo{year}{2017}).

\bibitem{Ning_JMPS2018}
\bibinfo{author}{Ning, X.} \& \bibinfo{author}{Pellegrino, S.}
\newblock \bibinfo{title}{Searching for imperfection insensitive externally
  pressurized near-spherical thin shells}.
\newblock \emph{\bibinfo{journal}{J. Mech. Phys. Solids}}
  \textbf{\bibinfo{volume}{120}}, \bibinfo{pages}{49--67}
  (\bibinfo{year}{2018}).

\bibitem{Gerasimidis_JAM2018}
\bibinfo{author}{Gerasimidis, S.}, \bibinfo{author}{Virot, E.},
  \bibinfo{author}{Hutchinson, J.~W.} \& \bibinfo{author}{Rubinstein, S.~M.}
\newblock \bibinfo{title}{On establishing buckling knockdowns for
  imperfection-sensitive shell structures}.
\newblock \emph{\bibinfo{journal}{Journal of Applied Mechanics}}
  \textbf{\bibinfo{volume}{85}}, \bibinfo{pages}{091010}
  (\bibinfo{year}{2018}).

\bibitem{Sieber_JAM2020}
\bibinfo{author}{Sieber, J.}, \bibinfo{author}{Hutchinson, J.~W.} \&
  \bibinfo{author}{Thompson, J. M.~T.}
\newblock \bibinfo{title}{Buckling thresholds for pre-loaded spherical shells
  subject to localized blasts}.
\newblock \emph{\bibinfo{journal}{J. Appl. Mech}}
  \textbf{\bibinfo{volume}{87}}, \bibinfo{pages}{031013}
  (\bibinfo{year}{2020}).

\bibitem{Yan_JMPS2020}
\bibinfo{author}{Yan, D.}, \bibinfo{author}{Pezzulla, M.} \&
  \bibinfo{author}{Reis, P.~M.}
\newblock \bibinfo{title}{Buckling of pressurized spherical shells containing a
  through-thickness defect}.
\newblock \emph{\bibinfo{journal}{J. Mech. Phys. Solids}}
  \textbf{\bibinfo{volume}{138}}, \bibinfo{pages}{103923}
  (\bibinfo{year}{2020}).

\bibitem{Virot_PRL2017}
\bibinfo{author}{Virot, E.}, \bibinfo{author}{Kreilos, T.},
  \bibinfo{author}{Schneider, T.~M.} \& \bibinfo{author}{Rubinstein, S.~M.}
\newblock \bibinfo{title}{Stability landscape of shell buckling}.
\newblock \emph{\bibinfo{journal}{Phys. Rev. Lett.}}
  \textbf{\bibinfo{volume}{119}}, \bibinfo{pages}{224101}
  (\bibinfo{year}{2017}).

\bibitem{Marthelot_JAM2017}
\bibinfo{author}{Marthelot, J.}, \bibinfo{author}{L\'opez~Jim\'enez, F.},
  \bibinfo{author}{Lee, A.}, \bibinfo{author}{Hutchinson, J.~W.} \&
  \bibinfo{author}{Reis, P.~M.}
\newblock \bibinfo{title}{Buckling of a pressurized hemispherical shell
  subjected to a probing force}.
\newblock \emph{\bibinfo{journal}{J. Appl. Mech.}}
  \textbf{\bibinfo{volume}{84}}, \bibinfo{pages}{121005}
  (\bibinfo{year}{2017}).

\bibitem{Thompson_IntJBifurcatChaos2017}
\bibinfo{author}{Thompson, J. M.~T.}, \bibinfo{author}{Hutchinson, J.~W.} \&
  \bibinfo{author}{Sieber, J.}
\newblock \bibinfo{title}{Probing shells against buckling: a nondestructive
  technique for laboratory testing}.
\newblock \emph{\bibinfo{journal}{Int. J. Bifurcat. Chaos}}
  \textbf{\bibinfo{volume}{27}}, \bibinfo{pages}{1730048}
  (\bibinfo{year}{2017}).

\bibitem{Lee_SoftMatter2019}
\bibinfo{author}{Lee, A.}, \bibinfo{author}{Yan, D.},
  \bibinfo{author}{Pezzulla, M.}, \bibinfo{author}{Holmes, D.~P.} \&
  \bibinfo{author}{Reis, P.~M.}
\newblock \bibinfo{title}{Evolution of critical buckling conditions in
  imperfect bilayer shells through residual swelling}.
\newblock \emph{\bibinfo{journal}{Soft Matter}} \bibinfo{pages}{6134--6144}
  (\bibinfo{year}{2019}).

\bibitem{Loukaides_IntJSmartNanoMater2014}
\bibinfo{author}{Loukaides, E.~G.}, \bibinfo{author}{Smoukov, S.~K.} \&
  \bibinfo{author}{Seffen, K.~A.}
\newblock \bibinfo{title}{Magnetic actuation and transition shapes of a
  bistable spherical cap}.
\newblock \emph{\bibinfo{journal}{Int. J. Smart Nano. Mater.}}
  \textbf{\bibinfo{volume}{5}}, \bibinfo{pages}{270--282}
  (\bibinfo{year}{2014}).

\bibitem{Seffen_SmartMaterStruct2016}
\bibinfo{author}{Seffen, K.~A.} \& \bibinfo{author}{Vidoli, S.}
\newblock \bibinfo{title}{Eversion of bistable shells under magnetic actuation:
  a model of nonlinear shapes}.
\newblock \emph{\bibinfo{journal}{Smart Mater. Struct.}}
  \textbf{\bibinfo{volume}{25}}, \bibinfo{pages}{065010}
  (\bibinfo{year}{2016}).

\bibitem{Lum_PNAS2016}
\bibinfo{author}{Lum, G.~Z.} \emph{et~al.}
\newblock \bibinfo{title}{Shape-programmable magnetic soft matter}.
\newblock \emph{\bibinfo{journal}{Proc. Natl. Acad. Sci. U.S.A.}}
  \textbf{\bibinfo{volume}{113}}, \bibinfo{pages}{E6007--E6015}
  (\bibinfo{year}{2016}).

\bibitem{Psarra_JMPS2019}
\bibinfo{author}{Psarra, E.}, \bibinfo{author}{Bodelot, L.} \&
  \bibinfo{author}{Danas, K.}
\newblock \bibinfo{title}{Wrinkling to crinkling transitions and curvature
  localization in a magnetoelastic film bonded to a non-magnetic substrate}.
\newblock \emph{\bibinfo{journal}{J. Mech. Phys. Solids}}
  \textbf{\bibinfo{volume}{133}}, \bibinfo{pages}{103734}
  (\bibinfo{year}{2019}).

\bibitem{Hu_Nature2018}
\bibinfo{author}{Hu, W.}, \bibinfo{author}{Lum, G.~Z.},
  \bibinfo{author}{Mastrangeli, M.} \& \bibinfo{author}{Sitti, M.}
\newblock \bibinfo{title}{Small-scale soft-bodied robot with multimodal
  locomotion}.
\newblock \emph{\bibinfo{journal}{Nature}} \textbf{\bibinfo{volume}{554}},
  \bibinfo{pages}{81--85} (\bibinfo{year}{2018}).

\bibitem{Kim_Nature2018}
\bibinfo{author}{Kim, Y.}, \bibinfo{author}{Yuk, H.}, \bibinfo{author}{Zhao,
  R.}, \bibinfo{author}{Chester, S.~A.} \& \bibinfo{author}{Zhao, X.}
\newblock \bibinfo{title}{Printing ferromagnetic domains for untethered
  fast-transforming soft materials}.
\newblock \emph{\bibinfo{journal}{Nature}} \textbf{\bibinfo{volume}{558}},
  \bibinfo{pages}{274} (\bibinfo{year}{2018}).

\bibitem{Kim_SciRobot2019}
\bibinfo{author}{Kim, Y.}, \bibinfo{author}{Parada, G.~A.},
  \bibinfo{author}{Liu, S.} \& \bibinfo{author}{Zhao, X.}
\newblock \bibinfo{title}{Ferromagnetic soft continuum robots}.
\newblock \emph{\bibinfo{journal}{Sci. Robot.}} \textbf{\bibinfo{volume}{4}},
  \bibinfo{pages}{eaax7329} (\bibinfo{year}{2019}).

\bibitem{Gu_NatCommun2020}
\bibinfo{author}{Gu, H.} \emph{et~al.}
\newblock \bibinfo{title}{Magnetic cilia carpets with programmable metachronal
  waves}.
\newblock \emph{\bibinfo{journal}{Nat. Commun.}} \textbf{\bibinfo{volume}{11}},
  \bibinfo{pages}{2637} (\bibinfo{year}{2020}).

\bibitem{Zhao_JMPS2019}
\bibinfo{author}{Zhao, R.}, \bibinfo{author}{Kim, Y.},
  \bibinfo{author}{Chester, S.~A.}, \bibinfo{author}{Sharma, P.} \&
  \bibinfo{author}{Zhao, X.}
\newblock \bibinfo{title}{Mechanics of hard-magnetic soft materials}.
\newblock \emph{\bibinfo{journal}{J. Mech. Phys. Solids}}
  \textbf{\bibinfo{volume}{124}}, \bibinfo{pages}{244--263}
  (\bibinfo{year}{2019}).

\bibitem{Gurtin_MechanicsThermodynamicsContinua2010}
\bibinfo{author}{Gurtin, M.~E.}, \bibinfo{author}{Fried, E.} \&
  \bibinfo{author}{Anand, L.}
\newblock \emph{\bibinfo{title}{The Mechanics and Thermodynamics of Continua}}
  (\bibinfo{publisher}{Cambridge University Press}, \bibinfo{year}{2010}).

\bibitem{Pezzulla_JAM2019}
\bibinfo{author}{Pezzulla, M.} \& \bibinfo{author}{Reis, P.~M.}
\newblock \bibinfo{title}{A weak form implementation of nonlinear axisymmetric
  shell equations with examples}.
\newblock \emph{\bibinfo{journal}{J. Appl. Mech.}}
  \textbf{\bibinfo{volume}{84}}, \bibinfo{pages}{034501}
  (\bibinfo{year}{2019}).

\bibitem{Bertotti_HysteresisMagnetism1998}
\bibinfo{author}{Bertotti, G.}
\newblock \emph{\bibinfo{title}{Hysteresis in Magnetism: for Physicists,
  Materials Scientists, and Engineers}} (\bibinfo{publisher}{Academic Press},
  \bibinfo{year}{1998}).

\bibitem{Pezzulla_PRL2018}
\bibinfo{author}{Pezzulla, M.}, \bibinfo{author}{Stoop, N.},
  \bibinfo{author}{Steranka, M.~P.}, \bibinfo{author}{Bade, A.~J.} \&
  \bibinfo{author}{Holmes, D.~P.}
\newblock \bibinfo{title}{Curvature-induced instabilities of shells}.
\newblock \emph{\bibinfo{journal}{Phys. Rev. Lett.}}
  \textbf{\bibinfo{volume}{120}}, \bibinfo{pages}{048002}
  (\bibinfo{year}{2018}).

\bibitem{Wang_JMPS2020}
\bibinfo{author}{Wang, L.}, \bibinfo{author}{Kim, Y.}, \bibinfo{author}{Guo,
  C.~F.} \& \bibinfo{author}{Zhao, X.}
\newblock \bibinfo{title}{Hard-magnetic elastica}.
\newblock \emph{\bibinfo{journal}{J. Mech. Phys. Solids}}
  \textbf{\bibinfo{volume}{142}}, \bibinfo{pages}{104045}
  (\bibinfo{year}{2020}).

\end{thebibliography}


\begin{thebibliography}{10}
\expandafter\ifx\csname url\endcsname\relax
  \def\url#1{\texttt{#1}}\fi
\expandafter\ifx\csname urlprefix\endcsname\relax\def\urlprefix{URL }\fi
\providecommand{\bibinfo}[2]{#2}
\providecommand{\eprint}[2][]{\url{#2}}

\bibitem{Lee_NatCommun2016}
\bibinfo{author}{Lee, A.} \emph{et~al.}
\newblock \bibinfo{title}{Fabrication of slender elastic shells by the coating
  of curved surfaces}.
\newblock \emph{\bibinfo{journal}{Nat. Commun.}} \textbf{\bibinfo{volume}{7}},
  \bibinfo{pages}{11155} (\bibinfo{year}{2016}).

\bibitem{Lee_JAM2016}
\bibinfo{author}{Lee, A.}, \bibinfo{author}{L\'opez~Jim\'enez, F.},
  \bibinfo{author}{Marthelot, J.}, \bibinfo{author}{Hutchinson, J.~W.} \&
  \bibinfo{author}{Reis, P.~M.}
\newblock \bibinfo{title}{The geometric role of precisely engineered
  imperfections on the critical buckling load of spherical elastic shells}.
\newblock \emph{\bibinfo{journal}{J. Appl. Mech.}}
  \textbf{\bibinfo{volume}{83}}, \bibinfo{pages}{111005}
  (\bibinfo{year}{2016}).

\bibitem{Timoshenko_TheoryPlatesShells1959}
\bibinfo{author}{Timoshenko, S.~P.} \& \bibinfo{author}{Woinowsky-Krieger, S.}
\newblock \emph{\bibinfo{title}{Theory of Plates and Shells}}
  (\bibinfo{publisher}{McGraw-Hill}, \bibinfo{year}{1959}).

\bibitem{Kim_Nature2018}
\bibinfo{author}{Kim, Y.}, \bibinfo{author}{Yuk, H.}, \bibinfo{author}{Zhao,
  R.}, \bibinfo{author}{Chester, S.~A.} \& \bibinfo{author}{Zhao, X.}
\newblock \bibinfo{title}{Printing ferromagnetic domains for untethered
  fast-transforming soft materials}.
\newblock \emph{\bibinfo{journal}{Nature}} \textbf{\bibinfo{volume}{558}},
  \bibinfo{pages}{274} (\bibinfo{year}{2018}).

\bibitem{Zhao_JMPS2019}
\bibinfo{author}{Zhao, R.}, \bibinfo{author}{Kim, Y.},
  \bibinfo{author}{Chester, S.~A.}, \bibinfo{author}{Sharma, P.} \&
  \bibinfo{author}{Zhao, X.}
\newblock \bibinfo{title}{Mechanics of hard-magnetic soft materials}.
\newblock \emph{\bibinfo{journal}{J. Mech. Phys. Solids}}
  \textbf{\bibinfo{volume}{124}}, \bibinfo{pages}{244--263}
  (\bibinfo{year}{2019}).

\bibitem{Bertotti_HysteresisMagnetism1998}
\bibinfo{author}{Bertotti, G.}
\newblock \emph{\bibinfo{title}{Hysteresis in Magnetism: for Physicists,
  Materials Scientists, and Engineers}} (\bibinfo{publisher}{Academic Press},
  \bibinfo{year}{1998}).

\bibitem{Niordson_ShellTheory1985}
\bibinfo{author}{Niordson, F.~I.}
\newblock \emph{\bibinfo{title}{Shell Theory}}.
\newblock North-Holland Series in Applied Mathematics and Mechanics
  (\bibinfo{publisher}{Elsevier Science}, \bibinfo{year}{1985}).

\bibitem{Ciarlet_TheoryShells2000}
\bibinfo{author}{Ciarlet, P.}
\newblock \emph{\bibinfo{title}{Theory of Shells}}.
\newblock Mathematical Elasticity (\bibinfo{publisher}{Elsevier Science},
  \bibinfo{year}{2000}).

\bibitem{Audoly_ElasticityGeometry2010}
\bibinfo{author}{Audoly, B.} \& \bibinfo{author}{Pomeau, Y.}
\newblock \emph{\bibinfo{title}{Elasticity and Geometry: from Hair Curls to the
  Non-linear Response of Shells}} (\bibinfo{publisher}{Oxford University
  Press}, \bibinfo{year}{2010}).

\bibitem{Pezzulla_JAM2019}
\bibinfo{author}{Pezzulla, M.} \& \bibinfo{author}{Reis, P.~M.}
\newblock \bibinfo{title}{A weak form implementation of nonlinear axisymmetric
  shell equations with examples}.
\newblock \emph{\bibinfo{journal}{J. Appl. Mech.}}
  \textbf{\bibinfo{volume}{84}}, \bibinfo{pages}{034501}
  (\bibinfo{year}{2019}).

\bibitem{Gurtin_MechanicsThermodynamicsContinua2010}
\bibinfo{author}{Gurtin, M.~E.}, \bibinfo{author}{Fried, E.} \&
  \bibinfo{author}{Anand, L.}
\newblock \emph{\bibinfo{title}{The Mechanics and Thermodynamics of Continua}}
  (\bibinfo{publisher}{Cambridge University Press}, \bibinfo{year}{2010}).

\bibitem{doCarmo_DifferentialGeometry2016}
\bibinfo{author}{do~Carmo, M.}
\newblock \emph{\bibinfo{title}{Differential Geometry of Curves and Surfaces:
  Revised and Updated Second Edition}}.
\newblock Dover Books on Mathematics (\bibinfo{publisher}{Dover Publications},
  \bibinfo{year}{2016}).

\bibitem{Oneill1997}
\bibinfo{author}{O'Neill, B.}
\newblock \emph{\bibinfo{title}{Elementary Differential Geometry}}
  (\bibinfo{publisher}{Academic Press}, \bibinfo{year}{1997}).

\bibitem{Koiter_NonlinearBuckling1969}
\bibinfo{author}{Koiter, W.~T.}
\newblock \bibinfo{title}{The nonlinear buckling behavior of a complete
  spherical shell under uniform external pressure, parts i, ii, iii \& iv}.
\newblock \emph{\bibinfo{journal}{Proc. Kon. Ned. Ak. Wet.}}
  \textbf{\bibinfo{volume}{B72}}, \bibinfo{pages}{40--123}
  (\bibinfo{year}{1969}).

\bibitem{Hutchinson_ProcMathPhysEngSci2016}
\bibinfo{author}{Hutchinson, J.~W.}
\newblock \bibinfo{title}{Buckling of spherical shells revisited}.
\newblock \emph{\bibinfo{journal}{Proc. Math. Phys. Eng. Sci.}}
  \textbf{\bibinfo{volume}{472}}, \bibinfo{pages}{20160577}
  (\bibinfo{year}{2016}).

\bibitem{Pezzulla_PRL2018}
\bibinfo{author}{Pezzulla, M.}, \bibinfo{author}{Stoop, N.},
  \bibinfo{author}{Steranka, M.~P.}, \bibinfo{author}{Bade, A.~J.} \&
  \bibinfo{author}{Holmes, D.~P.}
\newblock \bibinfo{title}{Curvature-induced instabilities of shells}.
\newblock \emph{\bibinfo{journal}{Phys. Rev. Lett.}}
  \textbf{\bibinfo{volume}{120}}, \bibinfo{pages}{048002}
  (\bibinfo{year}{2018}).

\bibitem{Hutchinson_JAM1967}
\bibinfo{author}{Hutchinson, J.~W.}
\newblock \bibinfo{title}{Imperfection sensitivity of externally pressurized
  spherical shells}.
\newblock \emph{\bibinfo{journal}{J. Appl. Mech.}}
  \textbf{\bibinfo{volume}{34}}, \bibinfo{pages}{49--55}
  (\bibinfo{year}{1967}).

\end{thebibliography}

\end{document}

\maketitle

\beginsupplement

\vspace{-1cm}
\section{Supplementary Notes on the Experimental Protocols}

In this section, we elaborate on the protocols that we have developed to fabricate and characterize our hard-magnetic shells. In Section~\ref{sec:shell_fabrication}, we present the coating technique used to fabricate the shell specimens. Then, in Section~\ref{sec:defect_characterization}, we discuss the methodology followed to characterize the geometry of the engineered defects of the fabricated shells.

\subsection{Fabrication protocol}
\label{sec:shell_fabrication}

\begin{figure}[h!]
\centering
\includegraphics[width=0.9\textwidth]{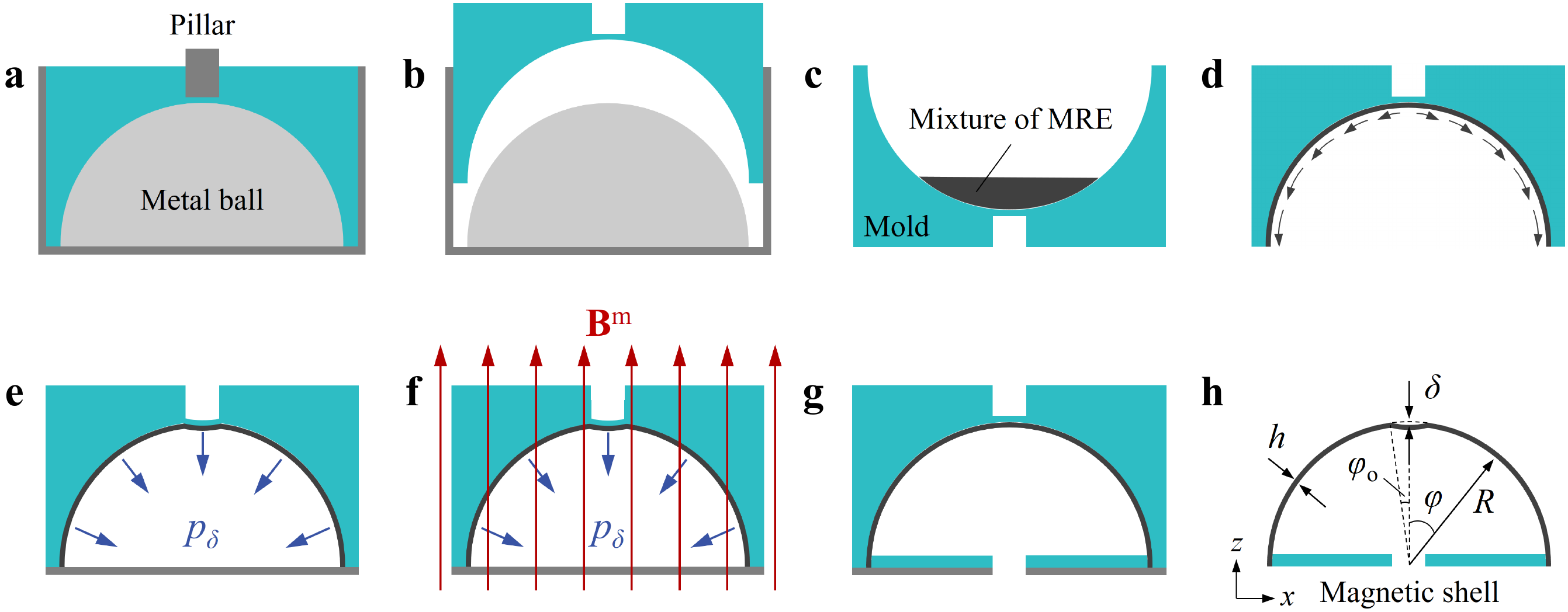}
\caption{\textbf{Schematic diagram of the fabrication protocol for precisely imperfect magnetic shells.} \textbf{a}, Fabrication of a negative spherical mold containing a soft spot (thin region) at its pole. \textbf{b}, Unmolding from the metal ball. \textbf{c}, The mold is filled with the liquid mixture of MRE. \textbf{d}, The mold is turned upside-down, which leads to a gravity-driven flow of the MRE that generates a thin coating on its surface. \textbf{e}, The soft spot (circular, thin, plate-like region) of the mold deforms under the applied negative pressure, $p_\delta$, to produce the geometric imperfection in the shell. \textbf{f}, The cured shell, while still adhered to the pressurized mold, is magnetized permanently using an applied magnetic field. \textbf{g}, A thick base is made at the equator by injection of additional VPS polymer from the base of the mold. \textbf{h}, The final magnetic shell containing a geometric defect is peeled off from the mold. The relevant geometric quantities are shown.} 
\label{fig:FigS1}
\end{figure}

We fabricated magnetic shell specimens using a customized coating protocol (see schematic in Fig.~\ref{fig:FigS1}) that was inspired from our previous work~\cite{Lee_NatCommun2016,Lee_JAM2016}, while also including some important new modifications. The protocol contains 3 steps -- (i) Fabrication of the mold, (ii) Fabrication of the shell, and (iii) Magnetization of the shell -- that are detailed next.
\begin{description}

\item[\textit{(i) Fabrication of the mold:}] An elastic mold was manufactured as the negative of a rigid hemisphere (radius $25.4\,$mm, stainless steel ball, TIS-GmbH) by pouring liquid VPS solution (base to catalyst ratio of 1:1 in weight, fully mixed and degassed) into a boxed-shaped container with the hemisphere placed on its bottom side (Fig.~\ref{fig:FigS1}\textbf{a}). A cylindrical pillar (radius $4.41\,$mm) was 3D-printed and positioned at a distance of $\approx0.5\,$mm above the north pole of the steel ball, in order to produce a circular thin region in the mold. After curing of the VPS, the mold was detached from both the ball and the pillar (Fig.~\ref{fig:FigS1}\textbf{b}). Then, we activated the surface of the mold using air plasma generated by a plasma cleaner (PDC-002-HPCE, Harrick Plasma) for 2 minutes and treated it with TMCS (Trimethylchlorosilane, Sigma-Aldrich) in a vacuum chamber for 1 hour. This treatment reduced the adhesion strength between the mold and the MRE to facilitate the demolding of shell specimens during step (ii). 

\item[\textit{(ii) Fabrication of the shell:}] The prepared mixture of MRE was poured into the negative spherical mold produced during step (i); see Fig.~\ref{fig:FigS1}\textbf{c} and the Methods section of the main article. The pouring was done after waiting for a set time, counting from the completion of mixture preparation ($100\,$s for MRE-22 and $20\,$s for MRE-32). The purpose of this waiting time was to increase the viscosity to a desired value~\cite{Lee_JAM2016}. With the spherical surface facing down, the gravity-driven viscous flow yielded a thin layer of MRE on the mold (Fig.~\ref{fig:FigS1}\textbf{d}). Before the MRE was fully cured, at about $13\,$min after pouring, we applied constant negative pressure to the mold by extracting the air inside using a syringe pump (Fig.~\ref{fig:FigS1}\textbf{e}). Under this depressurization, the soft spot (thin region) of the mold deformed, thereby producing an axisymmetric geometric imperfection in the shell at the pole. This defect was `frozen' in the shell after curing. During this depressurization stage, the bulk regions of the mold (other than the soft spot) exhibited negligible deformation. Shells made of MRE-22 or MRE-32 were fabricated using molds made out of the corresponding VPS polymer.

\item[\textit{(iii) Magnetization of the shell:}] After fabrication, the shells were then magnetized permanently by saturating the NdPrFeB particles in the MRE using an applied uniaxial magnetic field ($4.4\,$T, perpendicular to the equatorial plane, Fig.~\ref{fig:FigS1}\textbf{f}) generated by an impulse magnetizer (IM-K-010020-A, Magnet-Physik Dr. Steingroever GmbH). To constrain the shell in its natural undeformed configuration, the process was conducted before demolding by placing the cured shell (still adhered to the pressurized mold) inside the magnetization coil (MF-AsA-$\phi$70$\times$120mm, Magnet-Physik Dr. Steingroever GmbH). We then released the pressure loading and injected VPS mixture into the shell from its bottom to make a thick band ($\approx 6\,$mm in thickness, Fig.~\ref{fig:FigS1}\textbf{g}). This latter step ensured clamped boundary condition at the equator during buckling tests. Finally, the magnetic hemispherical shell containing a geometric defect was peeled off from the mold (Fig.~\ref{fig:FigS1}\textbf{h}). 
\end{description}

\captionsetup[figure]{labelfont={bf},name={Fig.}}

\begin{figure}[h]
\centering
\includegraphics[width=\textwidth]{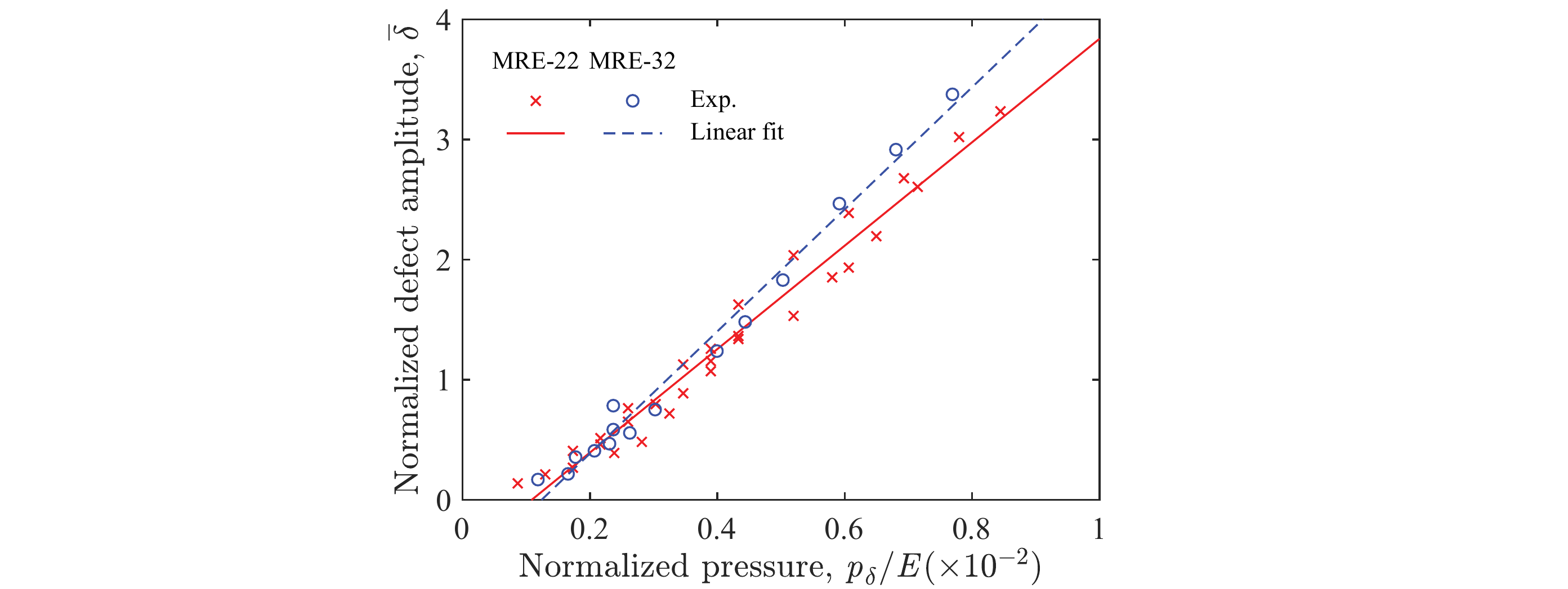}
\caption{ \textbf{Defect amplitude} normalized by the shell thickness, $\overline{\delta}=\delta/h$, versus pressure normalized by the Young's modulus of MRE, $p_\delta/E$, applied during the fabrication of MRE-22 and MRE-32 shells. Half angular width of the defects was fixed at $\varphi_\circ=11.7\,^\circ$ for MRE-22 shells and at $\varphi_\circ=11.6\,^\circ$ for MRE-32 shells. The experimental data are represented by the symbols and the linear fit by the solid (MRE-22) and dashed (MRE-32) lines.}
\label{fig:FigS2}
\end{figure}

\noindent \textbf{Systematic variation of the defect geometry.} The fabrication protocol presented above allows us to independently vary both the defect width (by changing the radius of the pillar used in the mold manufacture; Fig.~\ref{fig:FigS1}\textbf{a}) and the defect amplitude (by adjusting the pressure imposed on the mold; Fig.~\ref{fig:FigS1}\textbf{e}). For the fabrication of the MRE-22 shells, we applied pressure in the range of $1 \le p_\delta \le 9.75\,$kPa, which yielded defect amplitudes ranging from $\overline{\delta}=0.14$ to $\overline{\delta}=3.2$ (Fig.~\ref{fig:FigS2}). 

Throughout the study, the half angular width of the defects was fixed at $\varphi_\circ=11.7\pm0.1\,^\circ$ by setting the radius of the soft spot of the mold at $4.41\,$mm (corresponding to the target half angular width of $10.0\,^\circ$ set by the radius of the pillar). The minor difference between the obtained value for half angular width and the targeted value can be attributed to small deformations during fabrication near the boundary of the spot, which was not ideally clamped by the surrounding soft bulk regions of the mold. For the MRE-32 shells, we varied the defect amplitude in the range $0.17 \leq \overline{\delta} \leq 3.4$ by changing the pressure in the range $2 \le p_\delta \le 13\,$kPa, with a fixed half angular width of $\varphi_\circ=11.6\pm0.1\,^\circ$. An empirical linear relationship was found between the defect amplitude and the applied pressure (the slope depends on the bending stiffness of the thin region of the mold). These results serve as a convenient set of design guidelines to fabricate the specimens.

\subsection{Characterization of the defect geometry}
\label{sec:defect_characterization}

\noindent \textbf{Measurement of the defect profile.} To characterize the defect geometry of each shell specimen, the 3D profile of the shell (outer surface) was measured using an optical profilometer (VR-3200, Keyence Corporation). A perfect spherical surface was fitted to the measured data in the region away from the defect located at the pole of the shell. The defect profile, $w_\delta$, was defined as the radial distance between the fitting spherical surface and the measured profile. Due to the axisymmetry of the shell, the 3D profile was averaged latitude-wise to obtain the 2D profile. \\

\noindent \textbf{A simplified plate model to describe the generated defect geometry.} We used a simple model to provide an analytical description of the geometry of the defect. Note that, during pressurization of the mold, the bulk region outside of the soft spot exhibits negligible deformation compared to the soft spot. Moreover, the spherical cap of this soft spot is sufficiently small (with respect to the radius of curvature of the shell), such that it can be regarded as a (nearly) flat plate. Hence, we isolate this thin region and model it as a boundary-clamped circular flat plate loaded by a uniform pressure. The defect profile is determined by the deflection of this plate under the pressure loading. The governing equation can be obtained from classic Kirchhoff-Love theory for an isotropic and homogeneous plate under pure bending~\cite{Timoshenko_TheoryPlatesShells1959},
\begin{equation}
\label{eq:Kirchhoff-Love}
\frac{\partial^{4} w_\delta}{\partial x^{4}}+2 \frac{\partial^{4} w_\delta}{\partial x^{2} \partial y^{2}}+\frac{\partial^{4} w_\delta}{\partial y^{4}}=0\,,
\end{equation}
where $w_\delta$ is the out-of-plane displacement of the plate, and $x$,$y$ are the two in-plane coordinates, respectively. Solving Eq.~\eqref{eq:Kirchhoff-Love} for the specific case of a circular plate of radius, $r_\circ$, and thickness, $t$, under uniform pressure loading, $p_\delta$, and clamped boundary conditions, yields the following expression for the plate deflection~\cite{Timoshenko_TheoryPlatesShells1959}
\begin{equation}
\label{eq:plate_deflection}
w_\delta(r)=-\frac{p_\delta {r_\circ^4}}{64 D}\left(1-\frac{r^{2}}{r_\circ^{2}}\right)^{2}\,,
\end{equation}
where $D={E t^{3}}/{12\left(1-\nu^{2}\right)}$ is the bending modulus of the plate, and $r=\sqrt{x^2+y^2}$ is the radial coordinate. Note that we are using $t$ to designate the thickness of the circular plate, so as not to confuse it with the thickness of shells, $h$, used elsewhere in the study. Recognizing the maximum deflection at the center of the circular plate, $w_\delta(0)=-{p_\delta {r_\circ^4}}/{(64 D)}$ as the defect amplitude $\delta$, and $r/{r_\circ}$ as  $\approx\varphi/\varphi_\circ$ (from geometry), the profile of the dimple-like defect, for $\varphi\le\varphi_\circ$, can be then written as
\begin{equation}
\label{eq:defect_profile}
{w_\delta(\varphi)}=-{\delta} \left(1-\frac{\varphi^{2}}{\varphi_\circ^{2}}\right)^{2}\,,
\end{equation}
where $\varphi_\circ$ is the half angular width of the defect, and $\varphi$ is the angular distance (polar angle) measured from the center of the defect. Outside of the defect ($\varphi>\varphi_\circ$), we have ${w_\delta}=0$. 

We determined the amplitude, $\delta$, and half angular width, $\varphi_\circ$, of the defect by fitting Eq.~\eqref{eq:defect_profile} to the experimentally measured defect profiles, finding excellent agreement between the two (see Fig.~2\textbf{b} of the main article). From Eq.~\eqref{eq:plate_deflection}, we also note that $\delta=w_\delta(0)$ scales linearly with $p_\delta$, which allows us to vary the defect amplitude $\delta$ during shell fabrication by adjusting the pressure applied to the mold. In Fig.~\ref{fig:FigS2}, we present the values of $\delta$ measured after demolding versus $p_\delta$ for a series of MRE-22 and MRE-32 shells, confirming this linear relationship. The slope depends on the bending stiffness of the thin region of the mold. The non-zero intercept on the horizontal axis is due to the natural deflection of the thin region, which might be induced by the pre-stresses developed in the mold during the curing of VPS polymer.

So far, we have provided a simplified theoretical model to aid in the characterization of the profile of the defects in our imperfect shells. This analysis provides an excellent descriptor for the functional form of the defect geometry, even if the underlying two parameters (amplitude and width of the defect) are determined through fitting. In addition to aiding in setting the design parameters for fabrication, the geometric profile of the imperfect shells provided by this model is used directly as an input for the shell model (Section~\ref{sec:theorysection}). In Section~\ref{sec:parametric_study}, we will investigate how the geometric profile affects the onset of buckling by independently varying the underlying geometric parameters in the numerics.

\section{Supplementary Notes on the Experiments}

In Section~\ref{sec:magnetic_field}, we provide information on the generation of the magnetic field in our experiments, including the coils setup with the simulations and experiments performed to characterize the generated field. Then, in Section~\ref{sec:buckling_test}, we detail the apparatus and procedure for the buckling tests, with a focus on the generation of the magnetic field.

\subsection{Generation and characterization of the magnetic field}
\label{sec:magnetic_field}

\noindent \textbf{Description of the Helmholtz coils used to generate the magnetic field.} Two customized multi-turn circular coils with a square cross-section were set up in the Helmholtz configuration to generate a uniaxial uniform magnetic field in their central region. Each coil was manufactured by winding an aluminum circular spool with an insulated magnet wire: 18 turns in the axial direction and 19 turns in the radial direction (Repelec Moteurs S.A.). The magnet wire (enamelled wire, Isomet AG) had a circular cross-section of diameter $1\,$mm for the copper core and thickness $0.094\,$mm for the outer insulation layer. The final dimensions of the coil were: $72\,$mm in inner diameter, $114\,$mm in outer diameter, and $21\,$mm in height. The two identical coils were set concentrically, spaced with an axial centre-to-centre distance of $46.5\,$mm. They were connected in series and powered by a DC power supply that provided a maximum current/power of $25\,$A/$1.5\,$kW (EA-z 9200-25 T, EA-Elektro-Automatik GmbH). The strength of magnetic field was varied by adjusting the current output from the power supply.\\

\noindent \textbf{Numerical simulation of the generated magnetic field.} We used the commercial package COMSOL Multiphysics (v5.2, COMSOL Inc.) to simulate and characterize the magnetic field generated by the coils described above. These simulations were conducted using the \textit{Magnetic Fields} interface embedded in COMSOL, based on Amp\`{e}re's Law. Given the axisymmetry of the system, we modeled the 2D cross-section of the two coils, encompassed by a circular region of air (relative permeability 1) that was significantly larger than the size of the system (radius ten times larger than that of the coils). In this setting, we only consider the copper parts (relative permeability 0.999994) with current flowing through; \textit{i.e.}, the aluminum housing of the wires, with a relative permeability of $\approx 1$, were neglected. Since the strength of the magnetic field is linearly proportional to the current, $I$, we normalized the simulations by setting $I=1\,$A. \\

\begin{figure}[h!]
\centering
\includegraphics[width=\textwidth]{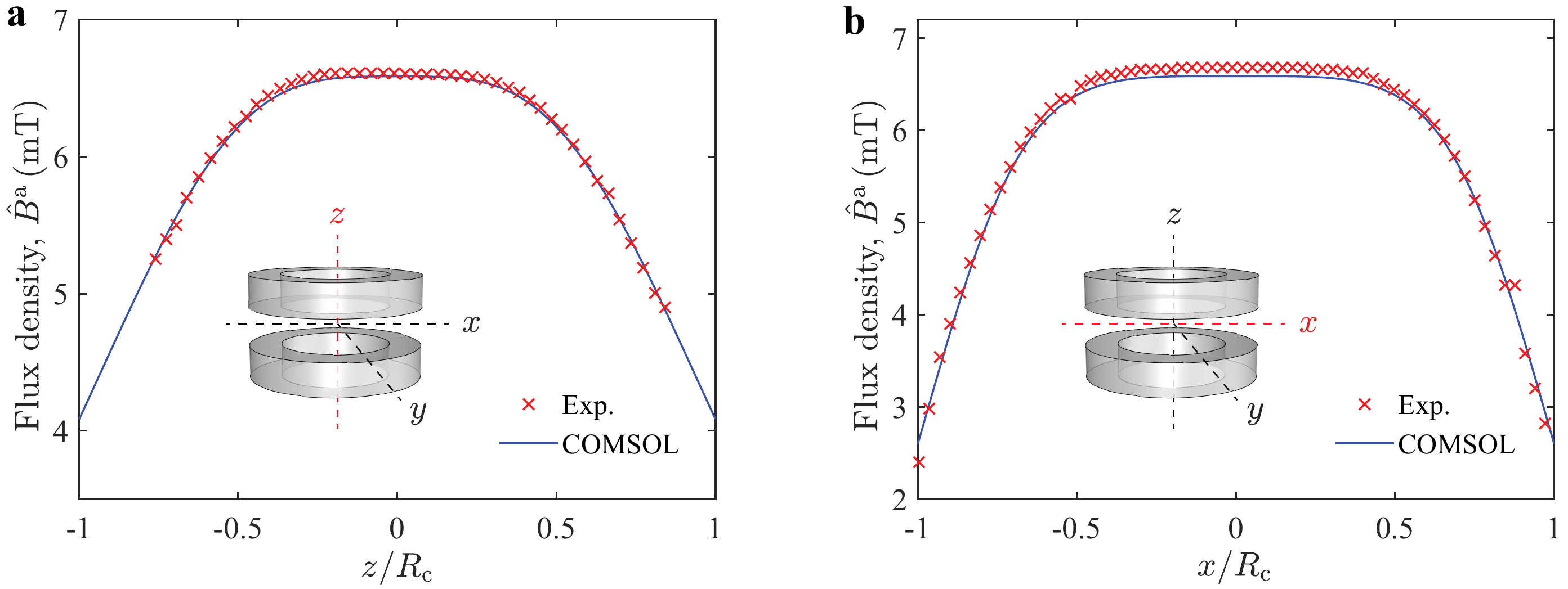}
\caption{\textbf{Comparison between the experimentally measured and the computed flux density of the generated magnetic field.} The experimental data (symbols) and simulation results obtained using COMSOL (solid lines) along the axial centre line ($x=y=0$) and the radial centre line ($y=z=0$) are presented in the plots \textbf{a} and \textbf{b}, respectively. A current of $1\,\mathrm{A}$ is applied to the coils. The radius of the coils is $R_\mathrm{c}$.}
\label{fig:FigS3}
\end{figure}

\noindent \textbf{Experimental characterization of the generated magnetic field.} Experimental measurements of the field strength were carried out to characterize the field generated in the experiments, as well as to validate the simulations. An axial Hall probe (HS-AGB5-4805, Magnet-Physik Dr. Steingroever GmbH) was moved along the central axis (vertical) of the two coils (inset of Fig.~\ref{fig:FigS3}\textbf{a}), and the corresponding flux density was measured by a Teslameter (FH 55, Magnet-Physik Dr. Steingroever GmbH). Similarly, a transverse Hall probe (HS-TGB5-104010, Magnet-Physik Dr. Steingroever GmbH) was used to measure the flux density along the radial (horizontal) direction (inset of Fig.~\ref{fig:FigS3}\textbf{a}). These measurements agree well with the simulation results, along both the axial and radial directions, as shown in Fig.~\ref{fig:FigS3}. 

\subsection{Buckling tests under combined magnetic and pressure loading}
\label{sec:buckling_test}

\noindent \textbf{Experimental apparatus for the buckling tests.} We measured the critical buckling load of the fabricated magnetic shells under pressure and magnetic loadings by positioning the shell in between the Helmholtz coils, with its pole located in the central region of the field (see Fig.~1\textbf{a},\textbf{b}). The pneumatic loading system comprised a syringe pump (NE-1000, New Era Pump Systems Inc.) and a differential pressure sensor (total error band $\pm3.7\,$Pa, 785-HSCDRRN002NDAA5, Honeywell International Inc.), which were both connected to the shell. \\

\noindent \textbf{Protocol of the buckling test.} For each experimental run, under a given external magnetic flux density, the syringe pump was set to extract the air inside the system at a constant flow rate of $0.4\,$mL/min, so as to depressurize the shell until it buckled. During depressurization, the pressure differential between the outside (atmospheric pressure) and inside of the shell was monitored by the pressure sensor at an acquisition rate of $8\,$Hz. For each tested shell, we obtained the evolution of pressure loading as a function of the imposed volume extracted by the syringe. The critical buckling pressure is defined as the peak value of this signal, representing the load-carrying capacity of the shell.

Avoiding overheating of the coils during the experimental tests was crucial. Otherwise, the resulting increase of temperature in the apparatus would have caused thermal expansion of air in the pneumatic loading system and compromised the results. This issue was particularly important for the stronger shells (at higher values of $B^\mathrm{a}$). Noting that the flow rate of $0.4\,$mL/min for the pressurization was set constant for all experiments, these experimental could run for up to 70\,s, which was far too long for the coils to be switched on. This increased running time for some of these runs meant that applying the strong magnetic field continuously throughout the full duration of the test was not feasible. Fortunately, auxiliary experiments demonstrated that the measured buckling pressure under the presence of the magnetic field was insensitive to when the field was switched on along the loading path, as long as this was done prior to buckling. Hence, to obtain reliable results, the field could be switched \textit{off} from the start of the experimental test and only switched \textit{on} in the proximity of $p_\mathrm{max}$, prior to reaching this value. To do so, we considered the cases of $B^\mathrm{a}>0$ (stronger shells with respect to no-field) and $B^\mathrm{a}<0$ (weaker shells with respect to no-field) differently. When the shell is strengthened by the magnetic field (tests for external flux density $B^\mathrm{a}>0$ in Figs.~2\textbf{c}~and~3 of the main article), buckling occurs at increasing pressure levels with $B^\mathrm{a}$.  To avoid overheating of the coils, the magnetic field was only applied when the pressure reached the level that was $10$\,Pa below the critical buckling pressure of the non-magnetic case ($B^\mathrm{a}=0$).  On the other hand, for shells weakened by the magnetic field (in the range $-66<B^\mathrm{a}\,\mathrm{[mT]}<0$ in Figs.~2\textbf{c}~and~3 of the main article), we followed a slightly different procedure while still attempting to minimize overheating. Specifically, we first measured the critical pressure for the lowest value considered for the applied magnetic field ($B^\mathrm{a}=-66$\,mT), corresponding to the weakest conditions (lowest buckling pressure), while switching on the field at the start of the test. For subsequent experiments, we systematically increased the applied field in steps of $6\,$mT, over the range $-66<B^\mathrm{a}\,\mathrm{[mT]}<0$ (11 experiments in total). For the $i^\mathrm{th}$ experiment, we had the field switched off until the pressure reached the value of $(p_\mathrm{max}(i-1)-5)\,[\mathrm{Pa}]$, where $p_\mathrm{max}(i-1)$ was the critical pressure measured in the previous $(i-1)^{\mathrm{th}}$ experiment.

\section{Supplementary Notes on the Theoretical Model}
\label{sec:theorysection}

This section is dedicated to the theoretical model that we developed to describe the onset of buckling of axisymmetric hard-magnetic shells.
First, we describe the kinematics of axisymmetric shells, followed by the dimensional reduction of the underlying Helmholtz free energy (Section~\ref{sec:shell_model}). To gain insight into this nonlinear magnetic energy, we perform an expansion for small displacements and interpret the physical meaning of the relevant terms at the different orders, namely: magnetic pressure, in-plane force, and torque  (Section~\ref{sec:physical_interpretation}). A scaling analysis on these terms is presented in Section~\ref{sec:scaling_analysis} to help rationalize how the knockdown factor changes with the applied magnetic field. In Section~\ref{sec:magnetic_effect}, we examine their role in the load-carrying behavior of magnetic shells. Emphasis is given to the second-order term of the magnetic energy representing the distributed magnetic torque relevant to the shell rotation, which is found to dominate the critical buckling conditions of the shells.

\subsection{Axisymmetric hard-magnetic shell model}

\label{sec:shell_model}

\noindent \textbf{Dimensional reduction of the Helmholtz free energy for axisymmetric shells.} As a starting point for the derivation of our hard-magnetic shell model, we use the Helmholtz free energy for ideal hard-magnetic soft materials~\cite{Kim_Nature2018,Zhao_JMPS2019}. This specialization of the free energy comprises an elastic part related to material deformation (hereon referred to as \textit{elastic energy}) and a magnetic part describing the work to align the residual magnetic moment along the external magnetic field (hereon referred to as \textit{magnetic energy}). Given that the saturated MRE can be assumed to have the permeability of vacuum~\cite{Bertotti_HysteresisMagnetism1998,Zhao_JMPS2019}, the magnetic potential energy in the reference configuration of volume $V$ can be simplified as the work to align the residual magnetic moment of the material with the external magnetic field. For 3D scale-free materials, this magnetic potential energy is~\cite{Bertotti_HysteresisMagnetism1998,Kim_Nature2018,Zhao_JMPS2019}
\begin{equation}
    \mathcal{U}_\mathrm{m}=-\int\frac{1}{\mu_0}\mathbf{F}{\mathbf{B}}^\mathrm{r}\cdot\mathbf{B}^\mathrm{a}\ \mathrm{d}V\,,
    \label{eq:magnetic_energy_3D}
\end{equation}
where $\mu_0$ is the permeability of vacuum, $\mathbf{F}$ is the deformation gradient tensor, ${\mathbf{B}}^\mathrm{r}$ is the vector of residual magnetic flux density of the material in the reference configuration, and $\mathbf{B}^\mathrm{a}$ is the vector of external magnetic flux density. 

In the following, we seek to derive the magnetic energy counterpart for axisymmetric shells by reducing the 3D description of Eq.~\eqref{eq:magnetic_energy_3D} to the 1D profile curve of the middle surface of the shell; a procedure that is commonly referred to as dimensional reduction~\cite{Niordson_ShellTheory1985, Ciarlet_TheoryShells2000,Audoly_ElasticityGeometry2010}. This dimensional reduction will be performed based on the Kirchhoff-Love assumptions for thin shells~\cite{Niordson_ShellTheory1985} (material fibers normal to the middle surface remain normal upon deformation without stretching nor shrinking), within the framework of small strains and large displacements/rotations. We will then follow a variational approach to minimize the total 1D energy via a weak form implementation, similarly to what we have recently proposed for elastic shells subjected to combined pressure and force-indentation loading~\cite{Pezzulla_JAM2019}.

\subsubsection{Kinematics of axisymmetric shells} 

Defining the spherical coordinates $(\varphi,\theta,\eta)$ that represent the polar angle, azimuthal angle, and radial distance, respectively, we parametrize the undeformed reference configuration of the shell as a solid body of revolution
\begin{equation}
\begin{aligned}
    \mathbf{X}(\varphi,\,\theta,\,\eta)&=\mathring{\mathbf{r}}(\varphi,\,\theta)+\eta\mathring{\mathbf{n}}(\varphi,\,\theta) \\ &=(\mathring{\rho}(\varphi)\cos\theta,\,\mathring{\rho}(\varphi)\sin\theta,\,\mathring{z}(\varphi))+\eta\mathring{\mathbf{n}}(\varphi,\,\theta)\,,
    \label{eq:undeformed_configuration}
\end{aligned}
\end{equation}
where $\mathring{\mathbf{r}}(\varphi,\theta)=(\mathring{\rho}(\varphi)\cos\theta,\,\mathring{\rho}(\varphi)\sin\theta,\,\mathring{z}(\varphi))$ represents the undeformed middle surface, with $(\mathring{\rho},\mathring{z})$ being the 2D Cartesian coordinates of its profile curve, and $\mathring{\mathbf{n}}(\varphi,\theta)$ is the undeformed unit normal vector. The covariant base vectors corresponding to $\mathbf{X}$ can be derived as
\begin{equation}
\begin{aligned}
    \mathbf{G}_1&=\frac{\partial \mathbf{X}}{\partial\varphi}=(\mathring{\rho}_{,\varphi}\cos\theta,\,\mathring{\rho}_{,\varphi}\sin\theta,\,\mathring{z}_{,\varphi})+\eta\mathring{\mathbf{n}}_{,\varphi}\,,\\
    \mathbf{G}_2&=\frac{\partial \mathbf{X}}{\partial\theta}=(-\mathring{\rho}\sin\theta,\,\mathring{\rho}\cos\theta,\,0)+\eta\mathring{\mathbf{n}}_{,\theta}\,,\\
    \mathbf{G}_3&=\frac{\partial \mathbf{X}}{\partial\eta}=\mathring{\mathbf{n}}=\frac{\mathring{\mathbf{r}}_{,\varphi}\times\mathring{\mathbf{r}}_{,\theta}}{|\mathring{\mathbf{r}},_\varphi\times\mathring{\mathbf{r}},_\theta|}=\frac{1}{\sqrt{\mathring{\rho},_\varphi^2+\mathring{z},_\varphi^2}}(-\mathring{z}_{,\varphi}\cos\theta,\,-\mathring{z}_{,\varphi}\sin\theta,\,\mathring{\rho}_{,\varphi})\,.
    \label{eq:undeformed_covariant}   
\end{aligned}
\end{equation}

Under the Kirchhoff-Love kinematic assumptions, the deformed configuration of the shell can be represented by the current middle surface ${\mathbf{r}}(\varphi,\theta)=({\rho}(\varphi)\cos\theta,{\rho}(\varphi)\sin\theta,{z}(\varphi))$, together with its normal vector $\mathbf{n}(\varphi,\,\theta)$, as
\begin{equation}
    \mathbf{x}(\varphi,\,\theta,\,\rho)=\mathbf{r}(\varphi,\,\theta)+\rho\mathbf{n}(\varphi,\,\theta)=({\rho}(\varphi)\cos\theta,\,{\rho}(\varphi)\sin\theta,\,{z}(\varphi))+\eta{\mathbf{n}}(\varphi,\,\theta)\,.
    \label{eq:deformed_configuration}
\end{equation}
The corresponding covariant base vectors are
\begin{equation}
\begin{aligned}
    \mathbf{g}_1&=\frac{\partial \mathbf{x}}{\partial\varphi}=({\rho}_{,\varphi}\cos\theta,\,{\rho}_{,\varphi}\sin\theta,\,{z},_\varphi)+\eta{\mathbf{n}}_{,\varphi}\,,\\
    \mathbf{g}_2&=\frac{\partial \mathbf{x}}{\partial\theta}=(-{\rho}\sin\theta,\,{\rho}\cos\theta,\,0)+\eta{\mathbf{n}},_\theta\,,\\
    \mathbf{g}_3&=\frac{\partial \mathbf{x}}{\partial\eta}={\mathbf{n}}=\frac{\mathbf{r}_\varphi\times\mathbf{r},_\theta}{|\mathbf{r},_\varphi\times\mathbf{r},_\theta|}=\frac{1}{\sqrt{{\rho},_\varphi^2+{z},_\varphi^2}}(-{z}_{,\varphi}\cos\theta,\,-{z}_{,\varphi}\sin\theta,\,{\rho}_{,\varphi})\,.
    \label{eq:deformed_covariant}   
\end{aligned}
\end{equation}
Now, we recall that the deformation gradient is a two-point tensor defined as~\cite{Gurtin_MechanicsThermodynamicsContinua2010}
\begin{equation}
   \mathbf{F}=\mathbf{g}_i\otimes\mathbf{G}^{i}=G^{ij}\mathbf{g}_i\otimes\mathbf{G}_j\,,
   \label{eq:F}  
\end{equation}
where $\mathbf{G}^{i}$, defined by $\mathbf{G}^{i}\cdot\mathbf{G}_{j}=\delta^{i}_{j}\,\,(i,j=1,2,3)$, are the contravariant base vectors corresponding to $\mathbf{X}$, and $G^{ij}$ are the contravariant components of the metric in the undeformed configuration of the shell. To compute $G^{ij}$, we first compute the covariant components $G_{ij}=\mathbf{G}_i\cdot\mathbf{G}_j$ as
\begin{equation}
\begin{aligned}
G_{11}&=\frac{\left(-\eta\mathring{z},_{\varphi\varphi} \mathring{\rho},_{\varphi}+\mathring{z},_{\varphi} \left(\eta \mathring{\rho},_{\varphi\varphi}+\mathring{z},_{\varphi}\sqrt{\mathring{z},_{\varphi}^2+\mathring{\rho},_{\varphi}^2}\right)+\mathring{\rho},_{\varphi}^2 \sqrt{\mathring{z},_{\varphi}^2+\mathring{\rho},_{\varphi}^2}\right)^2}{\bigl(\mathring{z},_{\varphi}^2+\mathring{\rho},_{\varphi}^2\bigr)^2}\,,\\
G_{22}&=\frac{\Bigl(\eta\mathring{z},_\varphi-\mathring{\rho}\sqrt{\mathring{\rho},_\varphi^2+\mathring{z},_\varphi^2}\Bigr)^2}{\mathring{\rho},_\varphi^2+\mathring{z},_\varphi^2}\,,\\
G_{33}&=1\,,\\
G_{12}&=0\,,\\
G_{13}&=0\,,\\
G_{23}&=0\,,
\label{eq:Gij_co} 
\end{aligned}
\end{equation}
resulting in a \textit{diagonal} metric of the undeformed shell configuration. Hence, the contravariant components can be readily computed as the inverse of the nonzero diagonal covariant components as
\begin{equation}
\begin{aligned}
G^{11}&=G_{11}^{-1}\,,\\
G^{22}&=G_{22}^{-1}\,,\\
G^{33}&=1\,,
\label{eq:Gij_contra} 
\end{aligned}
\end{equation}
with all the other components being zero. With these expressions at hand, we will expand the deformation gradient along the thickness coordinate so as to perform the dimensional reduction by integration of the magnetic energy along the thickness. Before doing so, we first need to briefly introduce the geometry of the middle surface, which will be later required for the development of our theoretical model.

\subsubsection{Geometry of the middle surface}

We obtain the parametrization of the undeformed middle surface, that is~$\mathring{\mathbf{r}}(\varphi,\theta)=(\mathring{\rho}(\varphi)\cos\theta,\,\mathring{\rho}(\varphi)\sin\theta,\,\mathring{z}(\varphi))$ by evaluating the parametrization of the undeformed three-dimensional shell~$\mathbf{X}$ for~$\eta=0$. 
The tangent vectors along the coordinate lines $\varphi$ and $\theta$ (\textit{i.e.}, the covariant base vectors of this surface) can be expressed, respectively, as
\begin{equation}
\begin{aligned}
    \mathring{\mathbf{a}}_1&=\frac{\partial \mathring{\mathbf{r}}}{\partial\varphi}=(\mathring{\rho}_{,\varphi}\cos\theta,\,\mathring{\rho}_{,\varphi}\sin\theta,\,\mathring{z}_{,\varphi})\,,\\
    \mathring{\mathbf{a}}_2&=\frac{\partial \mathring{\mathbf{r}}}{\partial\theta}=(-\mathring{\rho}\sin\theta,\,\mathring{\rho}\cos\theta,\,0)\,.
    \label{eq:undeformed_covariant_2D}   
\end{aligned}
\end{equation}
The corresponding contravariant base vectors $\mathring{\mathbf{a}}^{\alpha}$ are defined by $\mathring{\mathbf{a}}^{\alpha}\cdot\mathring{\mathbf{a}}_{\beta}=\delta^{\alpha}_{\beta}$ ($\alpha,\beta=1,2$). By definition~\cite{doCarmo_DifferentialGeometry2016}, the covariant components of the first and second fundamental forms of the undeformed middle surface can be expressed as $\mathring{a}_{\alpha\beta}=\mathring{\mathbf{a}}_{\alpha}\cdot\mathring{\mathbf{a}}_{\beta}$ and $\mathring{b}{_{\alpha\beta}}=-\mathring{\mathbf{n}},_\alpha \cdot \mathring{\mathbf{a}}_{\beta}$, respectively. Moreover, we denote the square root of the determinant of the covariant metric as~$\accentset{\circ}{a}=\sqrt{\det{(\accentset{\circ}{a}_{\alpha\beta})}}$, which will be used in the area measure.

Similarly, if we evaluate the parametrization of the deformed three-dimensional shell,~$\mathbf{x}$, at~$\eta=0$, we obtain the parametrization of the deformed middle surface, that is~$\mathbf{r}(\varphi,\theta)=(\rho(\varphi)\cos\theta,\,\rho(\varphi)\sin\theta,\,z(\varphi))$. 
The covariant base vectors can be expressed as
\begin{equation}
\begin{aligned}
    \mathbf{a}_1&=\frac{\partial \mathbf{r}}{\partial\varphi}=(\rho_{,\varphi}\cos\theta,\,\rho_{,\varphi}\sin\theta,\,z_{,\varphi})\,,\\
    \mathbf{a}_2&=\frac{\partial \mathbf{r}}{\partial\theta}=(-\rho\sin\theta,\,\rho\cos\theta,\,0)\,.
    \label{eq:deformed_covariant_2D}   
\end{aligned}
\end{equation}
The contravariant base vectors, first and second fundamental forms associated to the deformed middle surface can be computed with the same expressions we showed for the undeformed middle surface, employing the quantities without the accents. Having completed the geometric description of our system, we can now proceed to perform the dimensional reduction of the magnetic energy.

\subsubsection{Dimensional reduction of the magnetic energy for axisymmetric shells}

We seek to perform a dimensional reduction of the magnetic potential energy for 3D scale-free materials in Eq.~(\ref{eq:magnetic_energy_3D}), from 3D to 1D. To do so, consider the specific case for the configuration of the magnetic field in our experiments reported in the main article, where
\begin{equation}
\begin{aligned}
    {\mathbf{B}}^\mathrm{r}&={B}^\mathrm{r}\mathbf{e}_3\,,\\
    \mathbf{B}^\mathrm{a}&=B^\mathrm{a}\mathbf{e}_3\,.
    \label{eq:BrBa_exp} 
\end{aligned} 
\end{equation}

First of all, the elementary volume can be written as
\begin{equation}
    \textup{d}V=\mathring{a}(1-2\eta \mathring{H}+\eta^2\mathring{K})\,\mathrm{d}\varphi\mathrm{d}\theta\mathrm{d}\eta\,,
    \label{eq:dV} 
\end{equation}
where we used the expression of the square root of the determinant of the undeformed three-dimensional metric~$\mathring{G}$ for a stack of surfaces~\cite{Niordson_ShellTheory1985,Oneill1997}, namely~$\mathring{G}=\mathring{a}(1-2\eta \mathring{H}+\eta^2\mathring{K})$, with~$\mathring{H}$ and~$\mathring{K}$ denoting the mean and the Gaussian curvatures of the undeformed mid-surface, respectively. Then, expressing the deformation gradient by making use of Eq.~\eqref{eq:BrBa_exp}, \eqref{eq:Gij_co} and~\eqref{eq:Gij_contra}, we can write the integrand of Eq.~\eqref{eq:magnetic_energy_3D}, that is, the density of the magnetic potential, as
\begin{equation}
\begin{aligned}
    -\frac{1}{\mu_0}\mathbf{F}{\mathbf{B}}^\mathrm{r}\cdot\mathbf{B}^\mathrm{a}&=-\frac{1}{\mu_0}{B}^\mathrm{r}{B}^\mathrm{a}\mathbf{F}\mathbf{e}_3\cdot\mathbf{e}_3\,\\&=-\frac{1}{\mu_0}{B}^\mathrm{r}{B}^\mathrm{a}G^{ij}(\mathbf{g}_i\cdot\mathbf{e}_3)(\mathbf{G}_j\cdot\mathbf{e}_3)\,\\
    &=-\frac{1}{\mu_0}{B}^\mathrm{r}{B}^\mathrm{a}\left(G^{11}(\mathring{z},_\varphi+\eta\mathring{\mathbf{n}},_\varphi\cdot\mathbf{e}_3)(z,_\varphi+\eta\mathbf{n},_\varphi\cdot\mathbf{e}_3)+\frac{\mathring{\rho},_\varphi \rho,_\varphi}{\sqrt{\mathring{\rho},_\varphi^2+\mathring{z},_\varphi^2}\sqrt{\rho,_\varphi^2+z,_\varphi^2}}\right)\,.
    \label{eq:FBB} 
\end{aligned} 
\end{equation}

In the next step of the dimensional reduction procedure, we integrate the magnetic energy along the thickness of the reference configuration of the shell, by making use of Eqs.~(\ref{eq:dV}-\ref{eq:FBB}). The zeroth order terms in $\eta$ yield terms that are akin to a stretching energy (\textit{i.e.}, proportional to the thickness). The linear terms in $\eta$ vanish upon integration along the thickness direction from $\eta=-h/2$ to $\eta=h/2$. The higher-order terms, $\mathcal{O}(\eta^2)$, would give bending-like contributions to the energy, which we evaluated numerically to be much smaller than the stretching-like contribution, with no effect on the solution. Then, the integrand (Eq.~\eqref{eq:FBB}) becomes
\begin{equation}
    -\frac{1}{\mu_0}\mathbf{F}{\mathbf{B}}^\mathrm{r}\cdot\mathbf{B}^\mathrm{a}\,\textup{d}V=-\frac{1}{\mu_0}\left({\frac{\mathring{\rho},_\varphi \rho,_\varphi}{\sqrt{\mathring{\rho},_\varphi^2+\mathring{z},_\varphi^2}\sqrt{\rho,_\varphi^2+z,_\varphi^2}}+\frac{\mathring{z},_\varphi z,_\varphi}{\mathring{\rho},_\varphi^2+\mathring{z},_\varphi^2}}+\mathcal{O}(\eta^2)\right)\,\mathring{a}\mathrm{d}\varphi\mathrm{d}\theta\mathrm{d}\eta\,.
    \label{eq:FBB_0th} 
\end{equation}
Finally, we obtain the $1$D reduced magnetic energy for an axisymmetric shell
\begin{equation}
\begin{aligned}
    \mathcal{\overline{U}}_\mathrm{m}&=\frac{4(1-\nu^2)}{\pi EhR^2} \int_{-h/2}^{h/2}\int_0^{2\pi}\int_0^{\pi/2}-\frac{1}{\mu_0}\mathbf{F}{\mathbf{B}}^\mathrm{r}\cdot\mathbf{B}^\mathrm{a}\ \mathring{a}\mathrm{d}\varphi\mathrm{d}\theta\mathrm{d}\eta\,,\\
    &=-\frac{8(1-\nu^2)}{R^2}\frac{{B}^\mathrm{r}B^\mathrm{a}}{\mu_0 E}\int_0^{\pi/2}{\mathcal{F}(\mathring{\rho},_\varphi, \mathring{z},_\varphi, \rho,_\varphi, z,_\varphi)\ \mathring{a}\mathrm{d}\varphi}\,,
    \label{eq:magnetic_energy_1D} 
\end{aligned} 
\end{equation}
which we have normalized by~$\pi EhR^2/(4(1-\nu^2))$. For simplicity, we have introduced the following dimensionless function in Eq.~(\ref{eq:magnetic_energy_1D}) 
\begin{equation}
    \mathcal{F}=\frac{\mathring{\rho},_\varphi \rho,_\varphi}{\sqrt{\mathring{\rho},_\varphi^2+\mathring{z},_\varphi^2}\sqrt{\rho,_\varphi^2+z,_\varphi^2}}+\frac{\mathring{z},_\varphi z,_\varphi}{\mathring{\rho},_\varphi^2+\mathring{z},_\varphi^2},
    \label{eq:function_F} 
\end{equation}
which depends on the initial and deformed geometries of the shell. Also, note that  $\mathring{a}=\sqrt{\mathring{\rho}^2(\mathring{\rho},_\varphi^2+\mathring{z},_\varphi^2})$ is the area measure (\textit{i.e.}, the square root of the determinant of the metric of the undeformed middle surface). A dimensionless \textit{magneto-elastic parameter},
\begin{equation}
    \lambda_\mathrm{m}=\frac{{B}^\mathrm{r}B^\mathrm{a}}{\mu_0 E},
    \label{eq:magnetoelasticparameter}
\end{equation}
emerges from this analysis, representing the magneto-elastic coupling for thin magnetic shells.

\subsubsection{Elastic energy and live pressure potential}

So far, we have reduced the dimension of the magnetic potential from 3D to 1D, leaving the elastic energy and the live pressure potential aside as these are already known in the shell literature~\cite{Niordson_ShellTheory1985,Koiter_NonlinearBuckling1969,Hutchinson_ProcMathPhysEngSci2016,Pezzulla_JAM2019}. Next, we recall and briefly report these energies,  while emphasizing the underlying assumptions at their origin.

The elastic energy~$\mathcal{U}_\textup{e}$, as proposed by Koiter~\citep{Koiter_NonlinearBuckling1969,Niordson_ShellTheory1985}, is valid for small strains since it stems from the Kirchhoff-Saint Venant three-dimensional strain energy. Normalized by~$\pi EhR^2/(4(1-\nu^2))$, it reads~\cite{Pezzulla_JAM2019}

\begin{equation}\label{eq:elasticenergy}
\overline{\mathcal{U}}_\textup{e}=\frac{1}{R^2}\int_0^{\pi/2} [(1-\nu)\tr(\mathbf{a}-\accentset{\circ}{\mathbf{a}})^2+\nu\tr^2(\mathbf{a}-\accentset{\circ}{\mathbf{a}})]\accentset{\circ}{a} \,\textup{d} \varphi+\frac{1}{3}\left(\frac{h}{R}\right)^2\int_0^{\pi/2} [(1-\nu)\tr(\mathbf{b}-\accentset{\circ}{\mathbf{b}})^2+\nu\tr^2(\mathbf{b}-\accentset{\circ}{\mathbf{b}})]\accentset{\circ}{a} \,\textup{d} \varphi\,,
\end{equation}
where the first term is the (dimensionless) stretching energy, and the second term is the (dimensionless) bending energy. Note that we have introduced the trace operator in the undeformed metric ``$\tr$'' such that, for example,~$\tr(\mathbf{a})=\accentset{\circ}{a}^{\alpha\beta}a_{\alpha\beta}$.

The live pressure potential, normalized by~$\pi EhR^2/(4(1-\nu^2))$, can be expressed as~\cite{Pezzulla_JAM2019}
\begin{equation}
\overline{\mathcal{U}}_\textup{p}=8(1-\nu^2)\frac{p}{E}\Delta\overline{V}=8(1-\nu^2)\frac{1}{3hR^2}\frac{p}{E}\biggl[\int_0^{\pi/2}\mathbf{r}\cdot\mathbf{n}\,a\textup{d}\varphi-\int_0^{\pi/2}\accentset{\circ}{\mathbf{r}}\cdot\accentset{\circ}{\mathbf{n}}\,\accentset{\circ}{a}\textup{d}\varphi\biggr]\,.
\end{equation}

Finally, the total dimensionless energy is the sum of the elastic energy, the live pressure potential, and the magnetic energy:
\begin{equation}\label{eq:totalenergy}
\overline{\mathcal{U}}=\overline{\mathcal{U}}_\textup{e}+\overline{\mathcal{U}}_\textup{p}+\overline{\mathcal{U}}_\textup{m}\,.
\end{equation}\\

\noindent \textbf{Numerical implementation:}  In our computer simulations, the equilibrium equations can be obtained by minimizing the total energy~$\overline{\mathcal{U}}$ for all possible displacements of the shell. These equilibrium equations are solved via the Newton-Raphson method using the commercial package COMSOL Multiphysics (v. 5.2, COMSOL Inc.). Details of the implementation of our numerical procedure are provided in Ref.~\cite{Pezzulla_JAM2019}, albeit without the magnetic contribution derived above.

\subsection{Physical interpretation of the reduced magnetic energy}

\label{sec:physical_interpretation}

Above, we derived an expression for the reduced magnetic energy, $\mathcal{\overline{U}}_\mathrm{m}$ in Eq.~(\ref{eq:magnetic_energy_1D}). We now seek to provide a physical interpretation of the various terms involved in $\mathcal{\overline{U}}_\mathrm{m}$, focusing in the limit of small displacements. Due to axisymmetry, we express the displacement field of the shell profile in the covariant basis $\{\mathring{\mathbf{a}}_1,\mathring{\mathbf{n}}\}$, introduced before, as
\begin{equation}
    \mathbf{u}=\check{\mathbf{u}}+w\accentset{\circ}{\mathbf{n}}=u^\alpha\accentset{\circ}{\mathbf{a}}_\alpha+w\accentset{\circ}{\mathbf{n}}\,,
    \label{eq:displacement-field}
\end{equation}
with~$u^2=0$, because of axisymmetric deformations. Notice that~$u^{\alpha}$ are the contravariant components of the in-plane displacement field (along the covariant base vectors~$\mathring{a}_{\alpha}$ defined above), whereas~$w$ is the normal displacement~\cite{Niordson_ShellTheory1985}. Then, since~$\mathbf{r}=\mathring{\mathbf{r}}+\mathbf{u}$, we can expand the integrand of Eq.~\eqref{eq:magnetic_energy_1D} up to second order in the displacement field to find
\begin{equation}
\begin{aligned}
    \mathcal{\overline{U}}_\mathrm{m}=-\frac{8(1-\nu^2)}{R^2}\lambda_\mathrm{m}\int_0^{\pi/2}\Bigl(&1+\frac{\mathring{z},_\varphi^2}{(\mathring{\rho},_\varphi^2+\mathring{z},_\varphi^2)^2}\mathring{\mathbf{a}}_1\cdot\mathbf{u},_1-\frac{\mathring{\rho},_\varphi^2}{2(\mathring{\rho},_\varphi^2+\mathring{z},_\varphi^2)^2}\left(\mathring{\mathbf{n}}\cdot\mathbf{u},_1\right)^2\\
    &+\frac{\mathring{\rho},_\varphi\mathring{z},_\varphi}{(\mathring{\rho},_\varphi^2+\mathring{z},_\varphi^2)^{5/2}}\left(\mathring{\mathbf{n}}\cdot\mathbf{u},_1\right)\left(\mathring{\mathbf{a}}_1\cdot\mathbf{u},_1\right)\Bigr)\, \mathring{a}\mathrm{d}\varphi+\mathcal{O}(|\mathbf{u}|^3)\,,
    \label{eq:expansionsmalldisplacements}
\end{aligned}
\end{equation}
where the constant term, which does not depend on the displacement field, will be neglected from hereon.

We will now analyze each term in Eq.~(\ref{eq:expansionsmalldisplacements}), separately, with the goal of understanding their physical significance. Specifically, in the subsequent sub-sections, we will arrive at the following findings
\begin{itemize}[noitemsep]
\item[(i)] \textbf{The linear term} proportional to~$\mathring{\mathbf{a}}_1\cdot\mathbf{u},_1$, defined as 
\begin{equation}
        \mathcal{\overline{U}}_\mathrm{m}^{(1)}=-\frac{8(1-\nu^2)}{R^{2}}\lambda_\mathrm{m}\int_0^{\pi/2}\frac{\mathring{z},_\varphi^2}{(\mathring{\rho},_\varphi^2+\mathring{z},_\varphi^2)^2}\mathring{\mathbf{a}}_1\cdot\mathbf{u},_1\, \mathring{a} \mathrm{d}\varphi\,,
        \label{eq:firstorderterm}
\end{equation}
represents the combined action of distributed tangential and normal (dead pressure-like) forces;\\
\item[(ii)] \textbf{The second-order term} proportional to~$\left(\mathring{\mathbf{n}}\cdot\mathbf{u},_1\right)^2$, defined as
\begin{equation}
        \mathcal{\overline{U}}_\mathrm{m}^{(2)\text{A}}=\frac{8(1-\nu^2)}{R^{2}}\lambda_\mathrm{m}\int_0^{\pi/2}\frac{\mathring{\rho},_\varphi^2}{2(\mathring{\rho},_\varphi^2+\mathring{z},_\varphi^2)^2}\left(\mathring{\mathbf{n}}\cdot\mathbf{u},_1\right)^2\, \mathring{a} \mathrm{d}\varphi\,,
\end{equation}
is equivalent to a distribution of linear rotational springs, and dominates over all the other terms;\\
\item[(iii)] \textbf{The second-order term} proportional to~$\left(\mathring{\mathbf{n}}\cdot\mathbf{u},_1\right)\left(\mathring{\mathbf{a}}_1\cdot\mathbf{u},_1\right)$, which we define as
\begin{equation}
        \mathcal{\overline{U}}_\mathrm{m}^{(2)\text{B}}=-\frac{8(1-\nu^2)}{R^{2}}\lambda_\mathrm{m}\int_0^{\pi/2}\frac{\mathring{\rho},_\varphi\mathring{z},_\varphi}{(\mathring{\rho},_\varphi^2+\mathring{z},_\varphi^2)^{5/2}}\left(\mathring{\mathbf{n}}\cdot\mathbf{u},_1\right)\left(\mathring{\mathbf{a}}_1\cdot\mathbf{u},_1\right)\, \mathring{a} \mathrm{d}\varphi\,,
\end{equation}
represents a distribution of torques that are linear functions of the displacement field.
\end{itemize}

\subsubsection{First-order term in Eq.~(\ref{eq:expansionsmalldisplacements})} 

We start by considering the first-order term in Eq.~\eqref{eq:expansionsmalldisplacements}, defined as~$\mathcal{\overline{U}}_\mathrm{m}^{(1)}$ in Eq.~(\ref{eq:firstorderterm}), which is proportional to~$\mathring{\mathbf{a}}_1\cdot\mathbf{u},_1$. In words, this term represents the component of the longitudinal derivative of the displacement vector along the first covariant base vector (again longitude). Note that, using the covariant basis, we can establish the following identity 
\begin{equation}
    \mathring{\mathbf{a}}_1\cdot\mathbf{u},_1=\mathring{\mathbf{a}}_1\cdot[\check{\mathbf{u}},_1+(w\mathring{\mathbf{n}}),_1]=\mathring{\mathbf{a}}_1\cdot[\check{\mathbf{u}},_1+w,_1\mathring{\mathbf{n}}-w\mathring{b}_1^1\mathring{\mathbf{a}}_1]=\mathring{\mathbf{a}}_1\cdot\check{\mathbf{u}},_1-\mathring{b}_{11}w\,,
\end{equation}
where we used the Weingarten formulae~\cite{doCarmo_DifferentialGeometry2016} to express~$\mathring{\mathbf{n}},_1=-\mathring{b}_1^1\mathring{\mathbf{a}}_1$, with~$\mathring{b}_1^1$ denoting the mixed component along the latitudinal coordinate of the second fundamental form of the undeformed middle surface. We already considered the orthogonality of the spherical coordinates and the fact that the coordinate lines are principal. Then, the approximation for small displacements of the first-order term of the magnetic energy can be written as
\begin{equation}
    \mathcal{\overline{U}}_\mathrm{m}^{(1)}=-\frac{8(1-\nu^2)}{R^{2}}\lambda_\mathrm{m}\int_0^{\pi/2}\mathring{\psi}(\varphi)\Bigl(\frac{\mathring{\mathbf{a}}_1\cdot\check{\mathbf{u}},_1-\mathring{b}_{11}w}{R^{2}}\Bigr)\, \mathring{a} \mathrm{d}\varphi=\mathcal{\overline{U}}_\mathrm{m}^{(1u)}+\mathcal{\overline{U}}_\mathrm{m}^{(1w)}\,,
    \label{eq:firstorder_magneticenergy}
\end{equation}
where we highlighted the two different parts of this term proportional to the in-plane displacement and the normal displacement, respectively, and where
\begin{equation}
\mathring{\psi}(\varphi)=R^2\frac{\mathring{z},_\varphi^2}{(\mathring{\rho},_\varphi^2+\mathring{z},_\varphi^2)^2}\,,
\end{equation}
is a known dimensionless function depending on the undeformed configuration of the shell. \\

\noindent \textbf{Magnetic pressure.} If we compare the term~$\mathcal{\overline{U}}_\mathrm{m}^{(1w)}$, proportional to the normal displacement, to the potential of a dead pressure, namely $\mathcal{\overline{U}}_\mathrm{p}^\mathrm{dead}=\frac{8(1-\nu^2)}{R^2}\int _0^{\pi/2}\frac{p}{E}\frac{w}{h}\,\mathring{a} \mathrm{d}\varphi$ (positive pressure for a shell under compression), we can define a dimensionless contribution to the magnetic loading (normalized by $E$)  that is analogous to a dead pressure,
\begin{equation}
    \frac{p_\mathrm{m}(\varphi)}{E}=\lambda_\mathrm{m}\frac{h}{R}\frac{\mathring{b}_{11}}{R}\mathring{\psi}(\varphi)\,.
    \label{eq:magnetic_pressure}
\end{equation}
For a perfect spherical shell with $\mathring{\rho}=R\sin \varphi$, $\mathring{z}=R\cos \varphi$, $\mathring{b}_{11}=-R$, and~$\mathring{a}=R^2\sin \varphi$, Eq.~\eqref{eq:magnetic_pressure} reduces to
\begin{equation}
    \frac{p_\mathrm{m}(\varphi)}{E}=\lambda_\mathrm{m}\frac{h}{R}\sin^2(\varphi)\,.
\end{equation}
This equivalent magnetic dead pressure varies across the shell, being zero at the north pole. Notice that the magnitude of this normal force is dictated by the magneto-elastic parameter, $\lambda_m$, defined in Eq.~(\ref{eq:magnetoelasticparameter}), with a spatial variation imposed via~$\mathring{\psi}(\varphi)$.\\

\noindent \textbf{Distributed, tangential magnetic forces.} Regarding the term proportional to the in-plane displacement, that is~$\mathcal{\overline{U}}_\mathrm{m}^{(1u)}$, we rewrite it as
\begin{equation}
\begin{aligned}
    \mathcal{\overline{U}}_\mathrm{m}^{(1u)}=&-\frac{8(1-\nu^2)}{R^2}\lambda_\mathrm{m}\int_0^{\pi/2} \mathring{\psi}(\varphi) \frac{\mathring{\mathbf{a}}_1\cdot\check{\mathbf{u}},_1}{R^{2}}\, \mathring{a}\mathrm{d}\varphi\,\\
    =&-\frac{8(1-\nu^2)}{R^{2}}\lambda_\mathrm{m}\int_0^{\pi/2}\mathring{\psi}(\varphi)\frac{\mathring{\mathbf{a}}_1\cdot\Bigl({u^1_{\,,1}}\mathring{\mathbf{a}}_1+u^1{\mathring{\mathbf{a}}_1},_1\Bigl)}{R^{2}}\, \mathring{a}\mathrm{d}\varphi\,\\
    =&-\frac{8(1-\nu^2)}{R^{2}}\lambda_\mathrm{m}\int_0^{\pi/2}\mathring{\psi}(\varphi)\frac{1}{R^2}\Bigl[{u_{1,1}}+\frac{1}{2}\frac{\mathring{a}_{11,1}}{\mathring{a}_{11}}{u_1}\Bigr]\, \mathring{a}\mathrm{d}\varphi\,,
    \label{eq:U1_force_derivation}
\end{aligned}
\end{equation}
where we have used the Gauss-Weingarten formula \cite{doCarmo_DifferentialGeometry2016}
\begin{equation}
    \mathring{\mathbf{a}}_{1,1}=\mathring{b}_{11}\mathring{\mathbf{n}}+\nabla_1{\mathbf{a}}_1=\mathring{b}_{11}\mathring{\mathbf{n}}+\mathring{\Gamma}^1_{11}\mathring{\mathbf{a}}_1=\mathring{b}_{11}\mathring{\mathbf{n}}+\frac{1}{2}\frac{\mathring{a}_{11,1}}{\mathring{a}_{11}}\mathring{\mathbf{a}}_1\,,
\end{equation}
with $\nabla_1$ denoting the covariant derivative along $\mathring{\mathbf{a}}_1$, and $\mathring{\Gamma}^1_{11}$ is a Christoffel symbol of the second kind~\cite{doCarmo_DifferentialGeometry2016}. Then we integrate Eq.~(\ref{eq:U1_force_derivation}) by parts to obtain
\begin{equation}
    \mathcal{\overline{U}}_\mathrm{m}^{(1u)}=-\frac{8(1-\nu^2)}{R^{4}}\lambda_\mathrm{m}\Bigl\{\mathring{\psi}(\varphi)\mathring{a}{u_{1}}\bigr|_0^{\pi/2} +\int_0^{\pi/2}\Bigl[-\bigl(\mathring{\psi}(\varphi)\mathring{a}\bigr)_{,1}{u_{1}}+\frac{1}{2}\mathring{\psi}(\varphi)\frac{\mathring{a}_{11,1}}{\mathring{a}_{11}}\mathring{a}{u_1}\Bigr]\, \mathrm{d}\varphi\Bigr\}\,.
    \label{eq:U1_force_derivation_final}
\end{equation}
The first addend on the RHS of Eq.~\eqref{eq:U1_force_derivation_final} vanishes because of the clamped boundary condition at the equator ($u_{1}|_{\varphi=\pi/2}=0$). Consequently,  Eq.~\eqref{eq:U1_force_derivation_final} becomes
\begin{equation}
    \mathcal{\overline{U}}_\mathrm{m}^{(1u)}=-\frac{8(1-\nu^2)}{R^2}\lambda_\mathrm{m}\int_0^{\pi/2}\frac{1}{R^2}\Bigl(-\frac{(\mathring{\psi}(\varphi)\mathring{a})_{,1}}{\mathring{a}}+\frac{1}{2}\mathring{\psi}(\varphi)\frac{\mathring{a}_{11,1}}{\mathring{a}_{11}}\Bigr){u_1}\, \mathring{a}\mathrm{d}\varphi\,,
\end{equation}
which we can also write as
\begin{equation}
    \mathcal{\overline{U}}_\mathrm{m}^{(1u)}=-\frac{8(1-\nu^2)}{R^{2}}\int_0^{\pi/2}\mathbf{f}_\textup{m}\cdot\check{\mathbf{u}}\, \mathring{a}\mathrm{d}\varphi\,,
\end{equation}
where we have defined the distributed tangential forces induced by the magnetic field as
\begin{equation}
    \mathbf{f}_\textup{m}=f^{\alpha}\mathring{\mathbf{a}}_{\alpha}=\frac{\lambda_\mathrm{m}}{R^2}\Bigl(-\frac{(\mathring{\psi}(\varphi)\mathring{a})_{,1}}{\mathring{a}}+\frac{1}{2}\mathring{\psi}(\varphi)\frac{\mathring{a}_{11,1}}{\mathring{a}_{11}}\Bigr)\mathring{\mathbf{a}}_1= f^1\mathring{\mathbf{a}}_1\,,\quad\mathrm{with}\quad f^2=0\,.
\end{equation}
In the case of a perfect spherical shell, this term is equal to
\begin{equation}
     f^1=-\frac{3}{2}\lambda_\mathrm{m}R^{-2}\sin(2\varphi)\,.
\end{equation}
In conclusion, the first-order contribution to the magnetic energy, 
\begin{equation}
    \mathcal{\overline{U}}_\mathrm{m}^{(1)}=\frac{8(1-\nu^2)}{R^{2}}\int_0^{\pi/2}\Bigl(\frac{p_\mathrm{m}}{E}\frac{w}{h}-\mathbf{f}_\textup{m}\cdot\check{\mathbf{u}}\Bigr)\,\mathring{a}\mathrm{d}\varphi\,,
    \label{eq:Um-1}
\end{equation}
can be regarded as a combination of distributed normal and tangential forces that vary across the shell. We now move forward to analyze the second-order terms of $\mathcal{\overline{U}}_\mathrm{m}$.

\subsubsection{Second-order terms in Eq.~(\ref{eq:expansionsmalldisplacements})}

\noindent \textbf{Magnetic torque proportional to the shell rotation.} We now turn to the term~$\mathcal{\overline{U}}_\mathrm{m}^{(2)\text{A}}$, which is the second-order term proportional to~$\left(\mathring{\mathbf{n}}\cdot\mathbf{u},_1\right)^2$. We can rewrite the term~$\mathring{\mathbf{n}}\cdot\mathbf{u},_1$ by using the covariant basis as
\begin{equation}
    \mathring{\mathbf{n}}\cdot\mathbf{u},_1=\mathring{\mathbf{n}}\cdot[(u^\alpha\mathring{\mathbf{a}}_\alpha),_1+(w\mathring{\mathbf{n}}),_1]=\mathring{\mathbf{n}}\cdot\Bigl[\Bigl({u^1_{\,,1}}+\frac{1}{2}\frac{\mathring{a}_{11,1}}{\mathring{a}_{11}}u^1-w\mathring{b}_1^1\Bigr)\mathring{\mathbf{a}}_1+(\mathring{b}_{11}u^1+w,_1)\mathring{\mathbf{n}}\Bigr]=w,_1+\mathring{b}{_1^1}u_1\,.
\end{equation}  
The quantity above is the first covariant component of the rotation vector~$\mathbf{q}=(\delta\mathring{\mathbf{a}}_{\alpha}\cdot\mathring{\mathbf{n}})\mathring{\mathbf{a}}^{\alpha}=-(\delta\mathring{\mathbf{n}}\cdot\mathring{\mathbf{a}}_{\alpha})\mathring{\mathbf{a}}^{\alpha}$, which is defined, for example in Ref.~\cite{Niordson_ShellTheory1985,Pezzulla_PRL2018}, such that the rotation of a material fiber defined by the unit vector~$\mathbf{t}$ is given by $\mathbf{q}\cdot\mathbf{t}$. The rotation vector for small displacement can be expressed as~\cite{Niordson_ShellTheory1985,Pezzulla_PRL2018}
\begin{equation}\label{eq:rotation}
    \mathbf{q}=(w,_{\alpha}+\mathring{b}{_{\alpha}^{\beta}}u_{\beta})\mathring{\mathbf{a}}^{\alpha}\,,
\end{equation} 
where $\mathring{b}{_{\alpha}^{\beta}}$ is the mixed components of the second fundamental form of the undeformed middle surface. In the case of axisymmetry, we have 
\begin{align}
    \mathbf{q}=q_1\mathring{\mathbf{a}}^1=(w,_{1}+\mathring{b}{_{1}^{1}}u_{1})\mathring{\mathbf{a}}^{1}\,,
\end{align}
with $q_2=0$. Therefore, we can write 
\begin{equation}
    \mathcal{\overline{U}}_\mathrm{m}^{(2)\text{A}}=\frac{8(1-\nu^2)}{R^2}\lambda_\mathrm{m}\int_0^{\pi/2}\frac{1}{2}\mathring{\chi}(\varphi)\mathbf{q}\cdot\mathbf{q}\,\mathring{a} \mathrm{d}\varphi=\frac{8(1-\nu^2)}{R^2}\int_0^{\pi/2}\frac{1}{2}k\mathbf{q}\cdot\mathbf{q}\,\mathring{a} \mathrm{d}\varphi\,,
    \label{eq:Um-21}
\end{equation}
where 
\begin{equation}
    k=\lambda_\mathrm{m}\mathring{\chi}(\varphi)\,,
    \label{eq:equivalentrotationstiffness}
\end{equation}
and
\begin{equation}
    \mathring{\chi}(\varphi)=\frac{\mathring{\rho},_\varphi^2}{\mathring{\rho},_\varphi^2+\mathring{z},_\varphi^2}\,
\end{equation}
is a dimensionless geometric function depending exclusively on the undeformed configuration of the shell, similarly to~$\mathring{\psi}(\varphi)$. In conclusion, the term~$\mathcal{\overline{U}}_\mathrm{m}^{(2)\text{A}}$ can be interpreted as the energy potential of a distribution of linear torques $\boldsymbol{\tau}$ along the shell, defined as
\begin{equation}
    \boldsymbol{\tau}=-k\mathbf{q}\,,
\end{equation}
where~$k$, defined in Eq.~(\ref{eq:equivalentrotationstiffness}), is the equivalent rotational stiffness. For a perfect hemispherical shell, $k$ is maximum at the pole and decreases monotonically to zero at the equator. \\

\noindent \textbf{Magnetic torque proportional to the in-plane displacement.} Finally, we turn into the second-order term proportional to~$\left(\mathring{\mathbf{n}}\cdot\mathbf{u},_1\right)\left(\mathring{\mathbf{a}}_1\cdot\mathbf{u},_1\right)$, that is~$\mathcal{\overline{U}}_\mathrm{m}^{(2)\text{B}}$, which can be written as
\begin{equation}
\begin{aligned}
    \mathcal{\overline{U}}_\mathrm{m}^{(2)\text{B}}&=-\frac{8(1-\nu^2)}{R^2}\lambda_\mathrm{m}\int_0^{\pi/2}\frac{\mathring{\rho},_\varphi\mathring{z},_\varphi}{(\mathring{\rho},_\varphi^2+\mathring{z},_\varphi^2)^{5/2}}\left(\mathring{\mathbf{n}}\cdot\mathbf{u},_1\right)\left(\mathring{\mathbf{a}}_1\cdot\mathbf{u},_1\right)\,\mathring{a} \mathrm{d}\varphi\,\\
    &=-\frac{8(1-\nu^2)}{R^2}\lambda_\mathrm{m}\int_0^{\pi/2}\frac{\mathring{\rho},_\varphi\mathring{z},_\varphi}{(\mathring{\rho},_\varphi^2+\mathring{z},_\varphi^2)^{5/2}}\left(w,_1+\mathring{b}{_1^1}u_1\right)\left[\Bigl(-\frac{(\mathring{\psi}(\varphi)\mathring{a})_{,1}}{\mathring{a}}+\frac{1}{2}\frac{\mathring{a}_{11,1}}{\mathring{a}_{11}}\mathring{\psi}(\varphi)\Bigr){u_1}-\mathring{b}_{11}w\right]\,\mathring{a} \mathrm{d}\varphi\,\\
    &=-\frac{8(1-\nu^2)}{R^2}\int_0^{\pi/2}\mathbf{m}(\mathbf{u})\cdot\mathbf{q}\,\mathring{a} \mathrm{d}\varphi\,,
    \label{eq:Um-22}
\end{aligned}
\end{equation}
where $\mathbf{m}(\mathbf{u})$ is a distributed torque, which is a linear function of the displacement field, expressed as
\begin{equation}
   \mathbf{m}(\mathbf{u})=m^{\alpha}\mathring{\mathbf{a}}_{\alpha}=\lambda_\mathrm{m}\frac{\mathring{\rho},_\varphi\mathring{z},_\varphi}{(\mathring{\rho},_\varphi^2+\mathring{z},_\varphi^2)^{3/2}}\left[\Bigl(-\frac{(\mathring{\psi}(\varphi)\mathring{a})_{,1}}{\mathring{a}}+\frac{1}{2}\frac{\mathring{a}_{11,1}}{\mathring{a}_{11}}\mathring{\psi}(\varphi)\Bigr){u_1}-\mathring{b}_{11}w\right]\mathring{\mathbf{a}}_1= m^1\mathring{\mathbf{a}}_1\,,\quad\mathrm{with}\quad m^2=0\,.
\end{equation}
Therefore, the term~$\mathcal{\overline{U}}_\mathrm{m}^{(2)\text{B}}$ corresponds to a distribution of torques across the shell, which depend linearly on the displacement field. We will analyze the magnitude of the first and second-order terms studied in this section and show how the term~$\mathcal{\overline{U}}_\mathrm{m}^{(2)\text{A}}$, corresponding to a distribution of linear rotational springs, dominates over the other terms.

\subsection{Scaling analysis}

\label{sec:scaling_analysis}

Now that we have a full interpretation of the magnetic energy -- Eqs.~\eqref{eq:Um-1}, \eqref{eq:Um-21} and~\eqref{eq:Um-22} --, we perform a scaling analysis to understand how the buckling pressure depends on the magnetic field, the geometry of the shell and its material properties. We will proceed by balancing the potential associated to the live pressure and the magnetic energy. 

First, we note that the undeformed surface coordinates, as well as their derivatives, scale as the characteristic radius of the shell; \textit{e.g.}, $\accentset{\circ}{\rho}\sim R$ and~$\accentset{\circ}{\rho},_{\varphi}\sim R$. Furthermore, as we are interested in the onset of buckling that takes place within the linear deformation regime, we assume that both the in-plane and normal displacements scale linearly with the shell thickness, $|\check{\mathbf{u}}|\sim h$ and $w\sim h$, all the way up to the onset of buckling.\\

\noindent \textbf{Scaling of the first-order term, ${\overline{\mathcal{U}}}_\textup{m}^{(1)}$.} Considering the above assumptions for the displacement field, we find the following scaling for the first-order term of the reduced magnetic energy in in Eq.~\eqref{eq:Um-1}:
\begin{equation}
   \overline{\mathcal{U}}_\textup{m}^{(1)}=\frac{8(1-\nu^2)}{R^2}\lambda_\textup{m}\int_0^{\pi/2} R^{-2}\Bigl[\accentset{\circ}{\psi}(\varphi)\accentset{\circ}{b}_{11}w-\Bigl(-\frac{(\accentset{\circ}{\psi}(\varphi)\accentset{\circ}{a})_{,1}}{\accentset{\circ}{a}}+\frac{1}{2}\accentset{\circ}{\psi}(\varphi)\frac{\accentset{\circ}{a}_{11,1}}{\accentset{\circ}{a}_{11}}\Bigr){u_1}\Bigr]\,\accentset{\circ}{a}\textup{d}\varphi\sim\lambda_\textup{m}\frac{h}{R}\,,
\end{equation}
noticing a linear proportionality to the thickness-to-radius ratio.
\\

\noindent \textbf{Scaling of the second-order term, ${\overline{\mathcal{U}}}_\textup{m}^{(2)\mathrm{A}}$.} We have to first focus on the scaling of the rotation vector~$\mathbf{q}$ in Eq.~\eqref{eq:rotation}. Given that the majority of the deformation of the shell takes place in the neighborhood of the north pole, within a length that scales with the boundary layer~$\sim\sqrt{Rh}$, angles will scale with the angular width of that boundary layer as ~$\sim\sqrt{h/R}$. This result means that~$|\mathbf{q}|\sim\sqrt{h/R}$ or, equivalently, $\mathbf{q}\cdot\mathbf{q}\sim h/R$. Finally, we can derive the scaling of~${\overline{\mathcal{U}}}_\textup{m}^{(2)\mathrm{A}}$ as
\begin{equation}
    {\overline{\mathcal{U}}}_\textup{m}^{(2)\mathrm{A}}=\frac{8(1-\nu^2)}{R^2}\lambda_\textup{m}\int_0^{\pi/2} \frac{1}{2} \accentset{\circ}{\chi}(\varphi) \mathbf{q}\cdot\mathbf{q}\,\accentset{\circ}{a}\textup{d}\varphi\sim\lambda_\textup{m}\frac{h}{R}\,,
\end{equation}
which scales exactly as the first-order term~${\overline{\mathcal{U}}}_\textup{m}^{(1)}$.\\

\noindent \textbf{Scaling of the second-order term, ${\overline{\mathcal{U}}}_\textup{m}^{(2)\mathrm{B}}$.} By applying the same arguments that we used for~${\overline{\mathcal{U}}}_\textup{m}^{(2)\mathrm{A}}$, the second-order term~${\overline{\mathcal{U}}}_\textup{m}^{(2)\mathrm{B}}$ scales as
\begin{equation}
  {\overline{\mathcal{U}}}_\textup{m}^{(2)\mathrm{B}}=-\frac{8(1-\nu^2)}{R^2}\int_0^{\pi/2}\mathbf{m}(\mathbf{u})\cdot\mathbf{q}\,\mathring{a} \mathrm{d}\varphi\sim \lambda_\textup{m}R^{-3}(\sqrt{Rh}+h)(Rh)\sim\lambda_\textup{m}\Bigl(\frac{h}{R}\Bigr)^{\frac{3}{2}}\,.
\end{equation}

\noindent \textbf{Scaling of the potential associated to the live pressure, $\overline{\mathcal{U}}_p$.} Since the change in volume scales as~$R^2h$, the dimensionless change in volume will be of order~$1$. Therefore, the live pressure potential scales as
\begin{equation}
    \overline{\mathcal{U}}_\textup{p}=8(1-\nu^2)\frac{p}{E}\Delta\overline{V}\sim\frac{p}{E}\,.
\end{equation}

\noindent \textbf{Scaling law of the change of knockdown factor,  $\Delta\kappa_\textup{d}$, with respect to the non-magnetic case.} We can now balance the live pressure potential with the reduced magnetic energy to obtain the scaling for the change of knockdown factor due to the applied magnetic field. Combining the scaling results for the first and second-order terms of the magnetic energy, we conclude that
\begin{equation}
  \overline{\mathcal{U}}_\textup{m}={\overline{\mathcal{U}}}_\textup{m}^{(1)}+{\overline{\mathcal{U}}}_\textup{m}^{(2)\mathrm{A}}+{\overline{\mathcal{U}}}_\textup{m}^{(2)\mathrm{B}}\sim\lambda_\textup{m}\frac{h}{R}.
  \label{eq:scaling_Um}
\end{equation}

Recalling Eq.~(1) of the main article, we note that the classic prediction for the buckling pressure of a perfect spherical shells scales as~$p_c\sim E(h/R)^2$. Hence, we find that the knockdown factor (defined as $\kappa_\mathrm{d}=p_\mathrm{max}/p_\mathrm{c}$, where $p_\mathrm{max}$ actual critical buckling pressure of the imperfect shell) scales as ~$\kappa_\textup{d}\sim p/E(R/h)^2$. Consequently, the change of knockdown factor with respect to the non-magnetic case, $\kappa_\textup{d}=\kappa_\mathrm{d}-\kappa_{\mathrm{d}}({{B}^\mathrm{a}=0)}$, scales as~$\Delta \kappa_\textup{d} \sim \Delta p/E(R/h)^2$. Balancing the scalings for the potential of $\Delta p$, $\overline{\mathcal{U}}_{\Delta p}=\Delta p/E$, and the magnetic energy in Eq.~\eqref{eq:scaling_Um} yields
\begin{equation}
    \frac{\Delta p}{E}\sim\lambda_\textup{m}\frac{h}{R}\,.
\end{equation}
Hence, the scaling of the change of knockdown factor is
\begin{equation}
    \Delta\kappa_\textup{d}\sim \frac{\Delta p}{E}\left(\frac{R}{h}\right)^2\sim\lambda_\textup{m}\frac{R}{h}\,,
    \label{eq:scaling_law_buckling}
\end{equation}
From the result in Eq.~(\ref{eq:scaling_law_buckling}), we define the \emph{magneto-elastic buckling parameter},
\begin{equation}
    \Lambda_\mathrm{m}=\lambda_\textup{m} \frac{R}{h},
    \label{eq:magnetoelasticbucklingparameter}
\end{equation}
which governs the knockdown factor under combined pressure and magnetic loading of spherical magnetic shells. The scaling $\Delta\kappa_\textup{d} \sim \Lambda_\mathrm{m}$ allows us to rationalize how the magnetic field modifies the buckling pressure for shells with different radius-to-thickness ratios. 

\subsection{Quantification of the effect of the magnetic field on shell buckling}
\label{sec:magnetic_effect}

Towards gaining a better physical understanding of the magnitude of the first and second order terms of the reduced magnetic energy, we ran two separate sets of simulations for shells containing a defect of amplitude~$\delta/h=0.39$ and angular width~$\overline{\varphi}_\circ=3.2$. First, we used (i) the fully nonlinear magnetic energy ($\overline{\mathcal{U}}_\mathrm{m}$, Eq.~\eqref{eq:magnetic_energy_1D}). Second, we considered its terms at the different orders: (ii) up to second-order ($\overline{\mathcal{U}}_\mathrm{m}^{(1)}+{\overline{\mathcal{U}}}_\textup{m}^{(2)\mathrm{A}}+{\overline{\mathcal{U}}}_\textup{m}^{(2)\mathrm{B}}$, Eqs.~\eqref{eq:Um-1},~\eqref{eq:Um-21}, and~\eqref{eq:Um-22}, respectively), (iii) the first-order term only ($\overline{\mathcal{U}}_\mathrm{m}^{(1)}$, Eq.~\eqref{eq:Um-1}), and (iv) the second-order term relevant to the shell rotation only (${\overline{\mathcal{U}}}_\textup{m}^{(2)\mathrm{A}}$, Eqs.~\eqref{eq:Um-21}). 

\begin{figure}[h!]
\centering
\includegraphics[width=\textwidth]{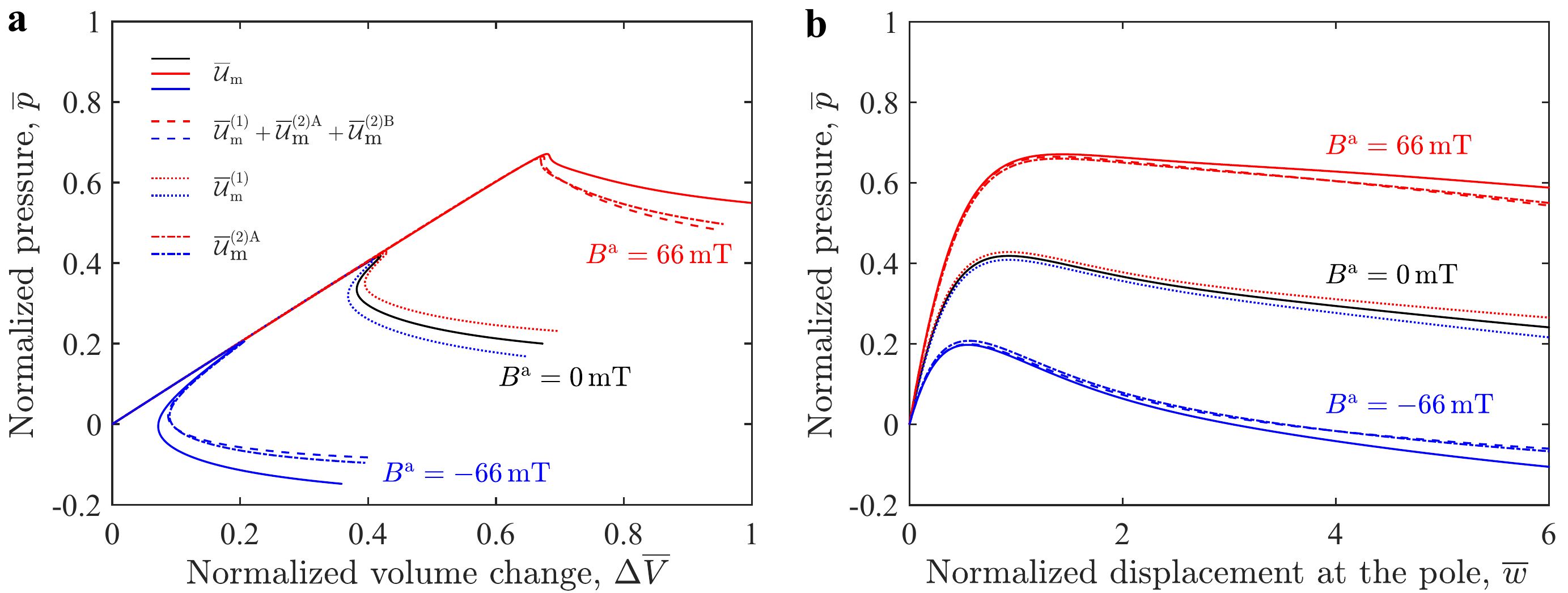}
\caption{\textbf{Load-carrying behavior predictions for a magnetic spherical shell from our 1D model.} Different terms of magnetic energy were considered in the model: (i) the fully nonlinear magnetic energy ($\overline{\mathcal{U}}_\mathrm{m}$, solid lines), (ii) the magnetic energy for small displacements up to second-order ($\overline{\mathcal{U}}_\mathrm{m}^{(1)}+{\overline{\mathcal{U}}}_\textup{m}^{(2)\mathrm{A}}+{\overline{\mathcal{U}}}_\textup{m}^{(2)\mathrm{B}}$, dashed lines), (iii) the first-order term only ($\overline{\mathcal{U}}_\mathrm{m}^{(1)}$, dotted lines), and (iv) the second-order term relevant to the shell rotation only (${\overline{\mathcal{U}}}_\textup{m}^{(2)\mathrm{A}}$, dash-dotted lines). The results are presented for different values of the external magnetic flux density ${B}^\mathrm{a}=\{66, 0, -66\}\,$mT, represented by the red, black, and blue curves, respectively. The shell contains a defect of amplitude $\delta/h=0.39$ and width $\overline{\varphi}_\circ=3.2$, which corresponds to the experiments in Fig.~1\textbf{d}. (a) Normalized pressure, $\overline{p}=p/p_\mathrm{c}$, as a function of normalized volume change, $\Delta\overline{V}=\Delta V/\Delta V_\mathrm{c}$, of the shell. Buckling occurs at the maximum value, which defines the knockdown factor. (b) Normalized pressure, $\overline{p}$, as a function of normalized displacement at the pole, $\overline{w}=w/h$.}
\label{fig:FigS4}
\end{figure}

In Fig.~\ref{fig:FigS4}, we plot the loading paths of magnetic shells in both the plane~$\bar{p}-\Delta\overline{V}$ (Fig.~\ref{fig:FigS4}\textbf{a}) and in the plane~$\bar{p}-\overline{w}$ (Fig.~\ref{fig:FigS4}\textbf{b}), for three different values of the applied magnetic field~${B}^\mathrm{a}=\{66, 0, -66\}\,$mT (red, black, and blue lines, respectively). In both plots, the black curves correspond to the classic case of pressure buckling of the shell ($B^{\mathrm{a}}=0\,$mT)~\cite{Lee_JAM2016}. The solid curves correspond to the results from the fully nonlinear magnetic energy ${\overline{\mathcal{U}}}_\textup{m}$. When the magnetic field is applied, the loading paths are modified with respect to the zero field case; there is a marked difference between the red/blue curves and the black curves. Consequently, there is a finite change of the knockdown factor, as reported in detail in Figs.~3~and~4 of the main article. The dotted lines correspond to the loading paths obtained considering only the linear term of the reduced magnetic energy; clearly showing that the linear term alone is unable to reproduce the observed finite change in knockdown factor. By contrast, we find that the second-order magnetic energy term~${\overline{\mathcal{U}}}_\textup{m}^{(2)\mathrm{A}}$ is able to accurately reproduce the fully nonlinear behavior; the dotted-dashed curves are close to the solid curves, especially in the pre-buckling regime, up to the onset of the instability. Therefore, the second-order term in the magnetic energy, which includes the contribution of the distributed magnetic torque relevant to the shell rotation, dominates the modification of the critical buckling conditions due to the applied magnetic field.\\

\noindent \textbf{Distribution of the magnetic torque and its dependence on the applied pressure.} The magnetic torque, $\boldsymbol{\tau}=-k(\varphi)\mathbf{q}$, is non-uniformly distributed along the shell, depending on the rotation vector, $\mathbf{q}$, and the equivalent rotational stiffness, $k(\varphi)=\lambda_\mathrm{m}\mathring{\chi}(\varphi)$ (Eq.~\eqref{eq:equivalentrotationstiffness}). Here, we want to understand how the magnetic torque varies as a function of the rotation vector, for positive and negative applied magnetic fields, so as to shed light on the mechanism that underlies the change in knockdown factor. To do so, we selected the representative shell containing a defect of amplitude $\overline{\delta}=0.39$ and width $\overline{\varphi}_\circ=3.2$, which was tested in the experiments (Fig.~1\textbf{d} of the main article). 

\begin{figure}[h!]
\centering
\includegraphics[width=\textwidth]{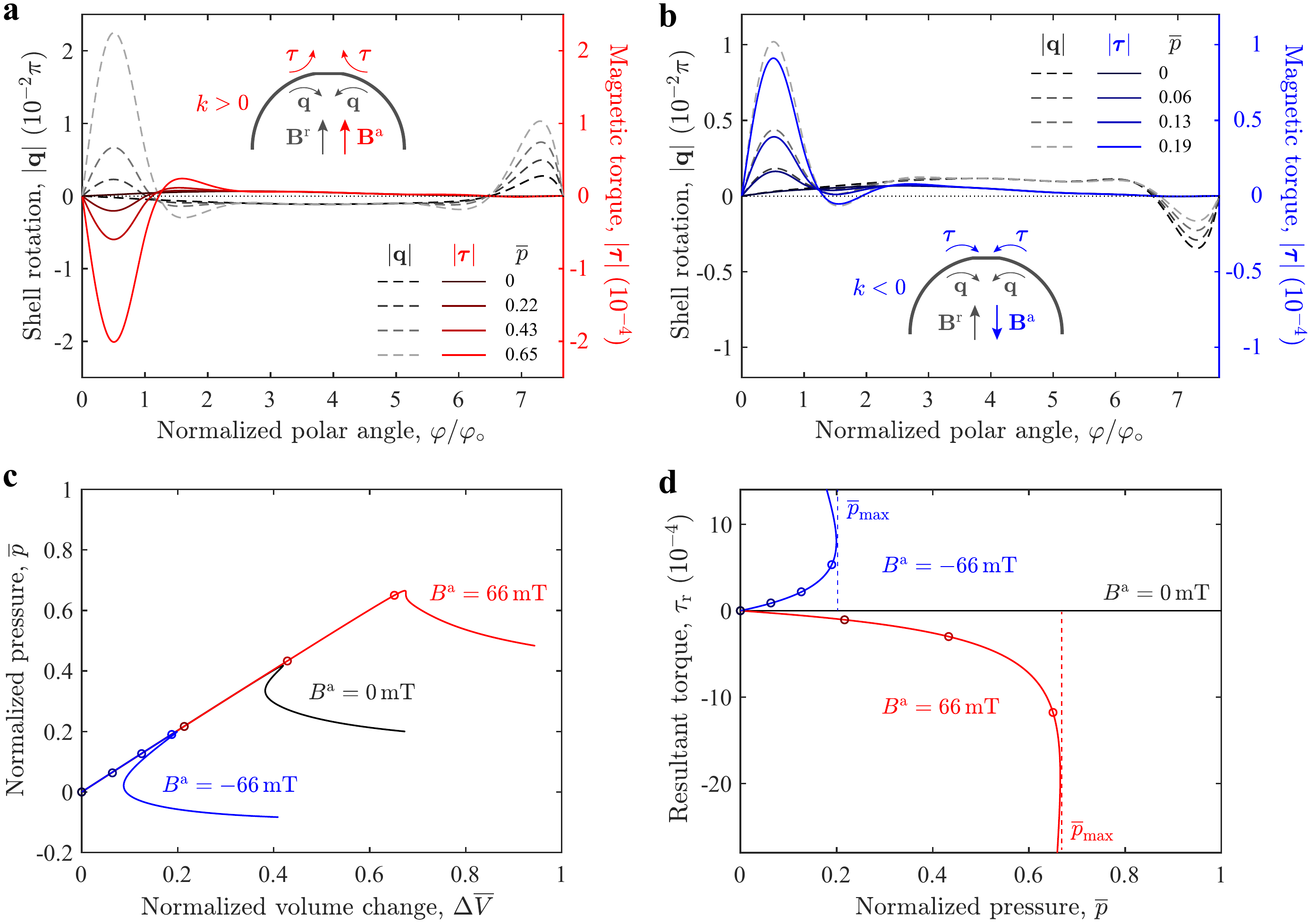}
\caption{\textbf{Shell rotation and magnetic torque predicted by our theoretical model.} Distributions of shell rotation  $|\mathbf{q}|=\sqrt{\mathring{a}^{11}}q_1$ (magnitude of $\mathbf{q}$) and of the associated magnetic torque $|\boldsymbol{\tau}|=\sqrt{\mathring{a}_{11}}\tau^1$ (magnitude of $\boldsymbol{\tau}$) along the shell 2D profile for $B^{\mathrm{a}}=66\,$mT (\textbf{a}) and $B^{\mathrm{a}}=-66\,$mT (\textbf{b}). The computations were carried out on the shell containing a defect of amplitude $\overline{\delta}=0.39$ and width $\overline{\varphi}_\circ=3.2$, which corresponds to the experiments in Fig.~1\textbf{d}. Terms up to second-order, $\overline{\mathcal{U}}_\mathrm{m}^{(1)}+{\overline{\mathcal{U}}}_\textup{m}^{(2)\mathrm{A}}+{\overline{\mathcal{U}}}_\textup{m}^{(2)\mathrm{B}}$, were considered for the magnetic energy. The results are presented at different values of the pressure level (see legend), corresponding to the open circles on the loading curves $\overline{p}(\Delta\overline{V})$ in \textbf{c} and $\tau_\mathrm{r}(\overline{p})$ curves in \textbf{d}. \textbf{d}, Normalized resultant magnetic torque imposed in the shell, $\tau_\mathrm{r}$, versus normalized pressure $\overline{p}$, corresponding to the three loading cases in \textbf{c}. The critical values, $\overline{p}_\mathrm{max}$, corresponds to the peaks of $\overline{p}(\Delta\overline{V})$ curves in \textbf{c}.} 
\label{fig:FigS5}
\end{figure}

In Fig.~\ref{fig:FigS5}, we plot the predictions from our model for the magnitudes of the rotation vector, $|\mathbf{q}|=\sqrt{\mathring{a}^{11}}q_1$, and the torque vector, $|\boldsymbol{\tau}|=\sqrt{\mathring{a}_{11}}\tau^1$, along the profile of the shell. Apart from the case of zero magnetic field, two representative values of the magnetic field were considered: $B^{\mathrm{a}}=66\,$mT (Fig.~\ref{fig:FigS5}\textbf{a}) and $B^{\mathrm{a}}=-66\,$mT (Fig.~\ref{fig:FigS5}\textbf{b}). For each case, we present results for the four pressure levels represented by the open circles marked on the loading curves in Fig.~\ref{fig:FigS5}\textbf{c}. We find that the shell rotation is localized both at the defect (\textit{i.e.}, at the pole of the shell, $\varphi=0$) and near the equator. Due to the non-uniformity of the equivalent rotational stiffness, $k$, the torque is prominent at the pole but almost zero in the rest of the shell. The magnetic torque increases with the rotation during pressurization. 

Note that, for the case when the field is parallel to the shell magnetization ($B^{\mathrm{a}}=66\,$mT), the directions of rotation and torque are opposite. By contrast, for the case when the field is anti-parallel (opposite) to the shell magnetization ($B^{\mathrm{a}}=-66\,$mT), the rotation and torque are in the same direction. We have integrated the distributed torque on the shell middle surface and, in Fig.~\ref{fig:FigS5}\textbf{d}, we plot the normalized resultant torque, $\tau_\mathrm{r}=\frac{1}{Rh}\iint{\sqrt{\mathring{a}_{11}}\tau}^1\ \mathring{a}\mathrm{d}\theta\mathrm{d}\varphi$, as a function of pressure. Note that the initial value of $\tau_\mathrm{r}$ at $p/p_\mathrm{c}=0$ has been subtracted. This initial value represents the total torque in the pressure-free state and was verified to have a negligible effect on the critical loads. Interestingly, we find that the torque does not increase linearly with pressure, with a slope that increases and becomes remarkably steep as the pressure approaches the critical value ($\overline{p}_\mathrm{max}$ represented by the dash lines in Fig.~\ref{fig:FigS5}\textbf{d}), where large rotation occurs.

In conclusion, we find that the change of the knockdown factor is dominantly induced by the torque localized near the defect, while the torque is almost zero throughout the rest of the shell, where rotations are negligible. As such, one should be able to extend the proposed approach, which has been demonstrated on hemispherical shells, for full spherical shells and even shallow shell sections, guaranteed by the localized nature of shell buckling. This would be an extension of classical methods in shell mechanics where, given the shallowness of the buckling mode, studies on the buckling of shallow spherical shells~\cite{Hutchinson_JAM1967} have proven to be successful in predicting the buckling of the corresponding full spherical shells~\cite{Koiter_NonlinearBuckling1969}.

\section{Parametric Study of the Effect of the Shell Geometry on Buckling}
\label{sec:parametric_study}

In this final section, we report the results from a parametric study where we have systematically varied the following two parameters: the shell radius-to-thickness ratio, $R/h$, and the defect width, $\overline{\varphi}_\circ$. We set out to investigate the effect of the shell geometry on buckling of magnetic shells. In Section~\ref{sec:radius_thickness_ratio}, we explored $R/h$ over a wider range ($50 \le R/h \le 400$) than that of the experiments ($R/h=79.1$ for MRE-22 shells and $R/h=91.3$ for MRE-32 shells), so as to more unequivocally establish the independence of $R/h$ on both the knockdown factor and its change under the applied magnetic field. The effect of the defect width on the critical buckling conditions is also investigated in Section~\ref{sec:defect_width}. 

\subsection{Independence of shell radius-to-thickness ratio}
\label{sec:radius_thickness_ratio}

\begin{figure}[h!]
\centering
\includegraphics[width=\textwidth]{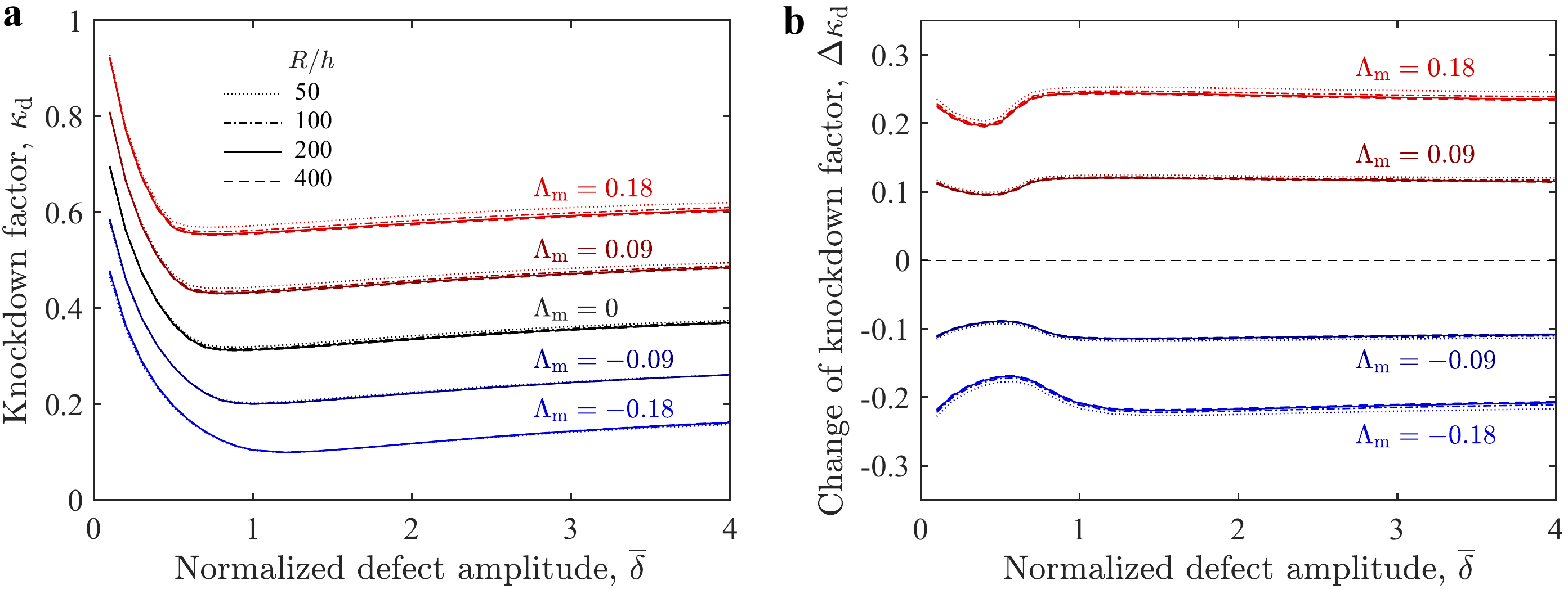}
\caption{\textbf{Independence of shell radius-to-thickness ratio.} \textbf{a}, Knockdown factor, $\kappa_\mathrm{d}$, and \textbf{b}, its change $\Delta\kappa_\mathrm{d}$ with respect to the non-magnetic case versus the amplitude of the defect, $\overline{\delta}$. The various values of the radius-to-thickness ratio are represented by the different line types: $R/h=50$ (dotted lines), $R/h=100$ (dash-dotted lines), $R/h=200$ (solid lines), and $R/h=400$ (dashed lines). The defect width is fixed at $\overline{\varphi}_\circ=3.2$. }
\label{fig:FigS6}
\end{figure}

Using the 1D model for magnetic shells that we presented in Section~\ref{sec:theorysection}, we obtained numerical solutions for a series of shells with different values of the radius-to-thickness ratio $R/h=\{50,100,200,400\}$, and quantified the predictions for the knockdown factor, $\kappa_\mathrm{d}$, as well as its change with respect to the non-magnetic case, $\Delta\kappa_\mathrm{d}$, of the shell. The results are shown in Fig.~\ref{fig:FigS6} for different values of the magneto-elastic buckling parameter $\Lambda_\mathrm{m}=\{0.18, 0.09, 0, -0.09, -0.18\}$. We recall from the definition in Eq.~\eqref{eq:magnetoelasticbucklingparameter} that $\Lambda_\mathrm{m}=\lambda_\mathrm{m} R/h$. We considered defects with increasing amplitudes $0.1 \le \overline{\delta} \le 4$, while fixing the defect width at $\overline{\varphi}_\circ=3.2$. We find that, for any given defect geometry, both $\kappa_\mathrm{d}$ and $\Delta\kappa_\mathrm{d}$ are governed by the parameter $\Lambda_\mathrm{m}$, independently of $R/h$ (there is a near-collapse of the $\kappa_\mathrm{d}(\overline{\delta})$ curves for the different $R/h$ values).

\subsection{Effect of width of the defect on the change of knockdown factor}

\label{sec:defect_width}

\begin{figure}[h!]
\centering
\includegraphics[width=\textwidth]{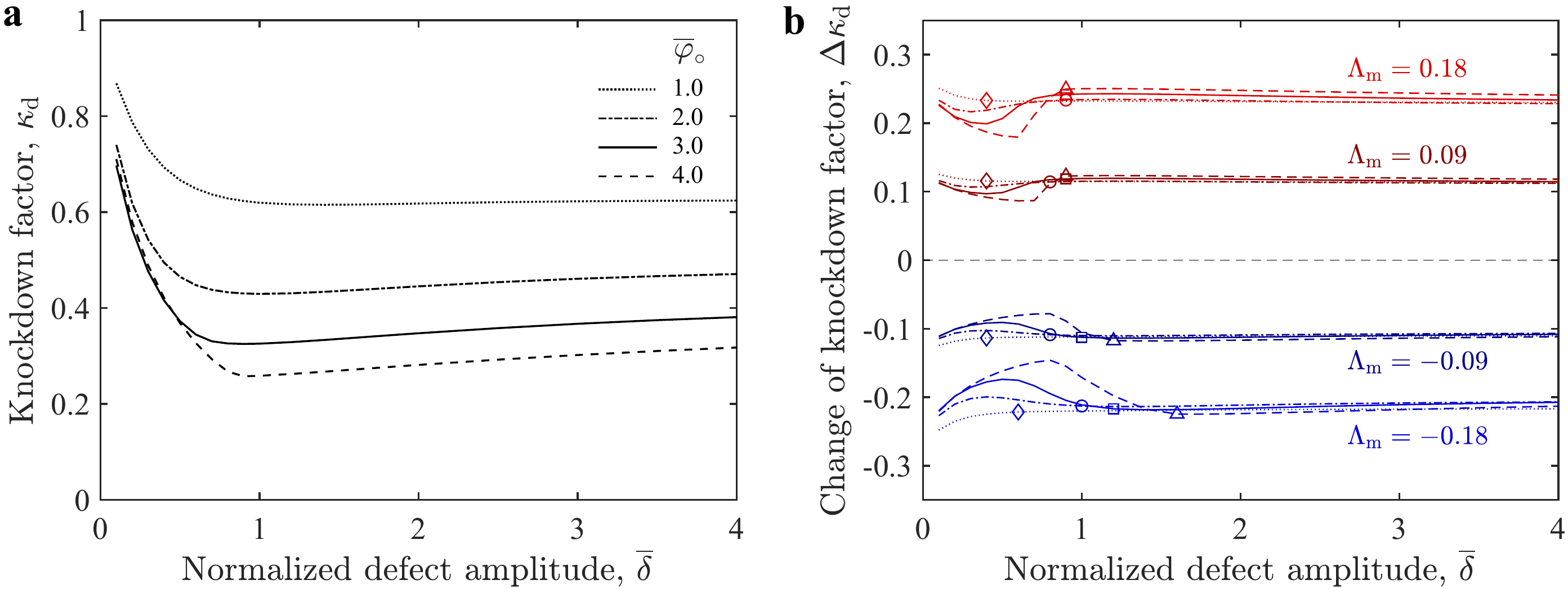}
\caption{\textbf{Effect of defect width.} \textbf{a}, Knockdown factor, $\kappa_\mathrm{d}$, versus defect amplitude, $\overline{\delta}$, for shells with different defect widths $\overline{\varphi}_\circ=\{1.0,2.0,3.0,4.0\}$ (as detailed in the legend), pressurized in the absence of the magnetic field. \textbf{b}, Change of knockdown factor, $\Delta\kappa_\mathrm{d}$, when imposing the magnetic field, versus defect amplitude, $\overline{\delta}$, for shells with different defect widths: $\overline{\varphi}_\circ=1.0$ (dotted lines), $\overline{\varphi}_\circ=2.0$ (dash-dotted lines), $\overline{\varphi}_\circ=3.0$ (solid lines), and $\overline{\varphi}_\circ=4.0$ (dashed lines). The external flux density is varied by $\Lambda_\mathrm{m}=\{0.18, 0.09, 0, -0.09, -0.18\}$. The shell radius-to-thickness ratio is chosen to be $R/h=200$. }
\label{fig:FigS7}
\end{figure}

We also investigated the effect of the defect width, $\overline{\varphi}_\circ$, on the critical buckling load using our magnetic shell model. We explored the following values $\overline{\varphi}_\circ=\{1.0,2.0,3.0,4.0\}$ and computed the prediction for both the  knockdown factor, $\kappa_\mathrm{d}$, of the shell and its change with respect to the non-magnetic case, $\Delta\kappa_\mathrm{d}$, for different magnetic loading conditions. In Fig.~\ref{fig:FigS7}\textbf{a}, we present the predicted $\kappa_\mathrm{d}(\overline{\delta})$ curves as a function of defect amplitude, $\overline{\delta}$, in the absence of the magnetic field ($\Lambda_\mathrm{m}=0$). We find that $\kappa_\mathrm{d}$ decreases for wider defects (\textit{i.e.}, increasing $\overline{\varphi}_\circ$). Then, we imposed the magnetic loading and varied the magnetic flux density, which is included in the magneto-elastic buckling parameter according to Eq.~\eqref{eq:scaling_law_buckling}, by setting the values: $\Lambda_\mathrm{m}=\{0.18, 0.09, 0, -0.09, -0.18\}$. The associated change of knockdown factor with respect to the non-magnetic case, $\Delta\kappa_\mathrm{d}$, is plotted in Fig.~\ref{fig:FigS7}\textbf{b}. We find that, when the defect amplitude is larger than a given threshold (determined by the differentiation of $\Delta\kappa_\mathrm{d}$ with respect to $\overline{\delta}$ smaller than 1\% and represented by the symbol on each curve of the plot) the change of knockdown factor becomes independent of the defect width (the curves for the various values of $\overline{\varphi}_\circ$ collapse). For relatively small defects ($\overline{\delta}$ below the threshold value), $\Delta\kappa_\mathrm{d}$ depends, but only slightly, on $\overline{\varphi}_\circ$. The corresponding variation is lower than 8\% for the extreme case explored of $\overline{\varphi}_\circ=4.0$ and $\Lambda_\mathrm{m}=-0.18$; less sensitively in compassion with $\kappa_\mathrm{d}$, which decreases by up to 45\% for the same parameter set.

\FloatBarrier
\newpage
\bibliographystyle{naturemag.bst}
\bibliography{references.bib}